\documentclass[a4paper,fleqn,usenatbib]{mnras}

\usepackage{newtxtext,newtxmath}

\usepackage[T1]{fontenc}
\usepackage{ae,aecompl}

\usepackage{graphicx}	
\usepackage{amsmath}	
\usepackage{amssymb}	

\usepackage{commath} 
\usepackage{caption,subcaption}
\usepackage{ulem} 
\renewcommand{\dif}{\mathop{}\!\mathrm{d}}
\newcommand{\bv}{\boldsymbol{\varv}}
\newcommand{\bx}{\mathbf x}

\graphicspath{%
	{./Figures/} %
}	

\title[Streaming model]{On the streaming model for redshift-space distortions
}

\author[J. Kuruvilla and C. Porciani]{
	Joseph Kuruvilla$^{1}$\thanks{E-mail: joseph@astro.uni-bonn.de}\thanks{Member of the International Max Planck Research School (IMPRS) for Astronomy and Astrophysics at the Universities of Bonn and Cologne} and
	Cristiano Porciani$^{1}$\thanks{E-mail: porciani@astro.uni-bonn.de}
	\\
	$^{1}$Argelander-Institut f\"ur Astronomie, Auf dem H\"ugel 71, D-53121 Bonn, Germany
}

\date{Accepted XXX. Received YYY; in original form ZZZ}

\pubyear{2018}

\begin{document}
	\label{firstpage}
	\pagerange{\pageref{firstpage}--\pageref{lastpage}}
	\maketitle
	

\begin{abstract}
The streaming model describes the mapping between real and redshift space for 2-point clustering statistics. Its key element is the probability density function (PDF) of line-of-sight pairwise peculiar velocities. Following a kinetic-theory approach, we derive the fundamental equations of the streaming model for ordered and unordered pairs. In the first case, we recover the classic equation while we demonstrate that modifications are necessary for unordered pairs. 
We then discuss several statistical properties of the pairwise velocities
for DM particles and haloes by using  a suite of high-resolution $N$-body simulations. We test the often used Gaussian ansatz for the PDF of pairwise velocities and discuss its limitations.
Finally, we introduce a mixture of Gaussians which is known in statistics as the generalised hyperbolic distribution and show that it provides an accurate fit to the PDF.
Once inserted in the streaming equation, the fit yields an excellent description of redshift-space correlations at all scales that vastly outperforms the Gaussian and exponential approximations. Using a principal-component analysis, we reduce the complexity of our model for large redshift-space separations. Our results 
increase the robustness of studies of anisotropic galaxy clustering and are useful for extending them towards smaller scales in order to test theories of gravity and interacting dark-energy models. 
\end{abstract}
	
	
	
	
	\section{Introduction}
	\label{sec:intro}
	
    Galaxy redshift surveys provide us with three-dimensional maps of the Universe. However, the resulting charts are twisted by the fact that we convert redshifts into distances by assuming a homogeneous model of cosmic expansion. It is well known that the measured redshift is not a perfect distance indicator in the presence of density perturbations.
Peculiar galaxy motions along the line of sight (los) generate the leading corrections and give rise to the so-called `redshift-space distortions' (RSD) between the reconstructed configuration (in `redshift space') and the actual galaxy distribution (in `real' space).

The clustering pattern in redshift space predominantly showcases two spurious features. Collapsed structures appear highly elongated along the los due to the velocity dispersion of their constituent galaxies,
a phenomenon known as the `Finger-of-God' (FoG) effect \citep{jackson72, sargentturner77}. 
At the same time, large-scale flows towards overdense regions (or away from underdense regions) coherently
deform the inferred galaxy distribution
\citep{kaiser87}. 
Overall, RSD break the rotational invariance of galaxy $N$-point statistics
which are anisotropic functions of the galaxy separations (in redshift space)
along the los and in the plane of the sky.
On large scales, the degree of anisotropy reflects the growth rate of cosmic structure and can thus be used to probe dark energy and gravity theories \citep[see][for a comprehensive review]{hamilton98}.

RSD are clearly detected in measurements of galaxy clustering
\citep[e.g.][]{DavisPeebles83,Fisher+94,Peacock+01,Guzzo+08} and
forthcoming surveys have the potential to extract valuable cosmological information from them \citep{WSP09,Giannantonio+,LIGER}.
The main limiting factor is modeling galaxy statistics in redshift space to the required accuracy for the widest possible range of scales.
This is a formidable problem involving four non-linear quantities:
the density and the velocity fields, galaxy biasing and the mapping from real to redshift space.
Several lines of research have been pursued over the last decades
with the aim of improving our understanding of RSD, 
including perturbative approaches for the largest scales \citep[e.g][]{Matsubara08,Taruya+09,TaruyaNi13,Senatore+14},
phenomenological models \citep[e.g.][]{Peacock-Dodds94}, and
combinations of the two \citep[e.g.][]{reid11}.

In this paper, we focus on the phenomenological approach introduced
by \citet{pe80} to model the galaxy two-point correlation function
in redshift space and subsequently generalised by \citet[][who dubbed it the `streaming model']{fi95}.
In this framework, the mapping from real to redshift
space is discussed in terms of the pairwise velocity distribution function.
Basically, (one plus) the anisotropic correlation function in redshift space is written as an integral
of (one plus) the spherically-symmetric correlation function in real space times the probability density of the relative los peculiar velocity. This model is exact in the distant-observer
limit. Its properties in Fourier space have been thoroughly investigated by \citet{sc04}. 

The history of the streaming model is rich and varied. 
In its early applications, it was used to model the small-scale clustering of galaxies from the CfA survey \citep{DavisPeebles83}. 
The pairwise velocity distribution
was assumed to be exponential\footnote{
Other analytical forms have also been considered but the exponential provided the best fit to the observational data.}. 
The (scale-dependent) mean pairwise (or streaming) velocity along the los was determined by requiring the conservation of galaxy pairs under the stable-clustering hypothesis while the corresponding velocity dispersion (assumed to be scale independent to first approximation) was treated as a free parameter and measured from the suppression of the redshift-space correlations along the los at fixed transverse separation
\citep[e.g.][]{DavisPeebles83, Bean83, Li+06}.
Later on,
\nocite{Fisher+94} 
\citet[][see also Fisher et al. 1994]{fi95} demonstrated that 
the streaming model with a Gaussian velocity distribution and
a scale-dependent velocity dispersion is compatible with
linear (Eulerian) perturbation theory on large scales. 
Therefore, it became clear that the character of the velocity PDF must change substantially 
with the galaxy separation, which makes the development of accurate theoretical models difficult.
For this reason,
hybrid models that combine linear perturbative predictions in redshift space with
a (scale-independent) streaming term for the incoherent small-scale motions
were introduced in order to describe galaxy clustering on a wider range of separations \citep{Peacock92,Park+94,Peacock-Dodds94}.
These `dispersion models' continue to be very popular \citep[e.g.][]{2dFHawkins+03,Guzzo+08,6dFBeutler+12,Chuang13}
although they correspond to a streaming model with a discontinous velocity PDF \citep{sc04}
and may lead to biased estimates of the cosmological parameters \citep{Kwan+12}.

Over time,
the `Gaussian streaming model' (GSM) has become the workhorse of RSD studies \citep{reid+12,Samushia+14,Alam+15,Chuang+17,Satpathy+17}. 
It requires three inputs: the real-space correlation function plus
the mean and the dispersion of the los relative velocity distribution
both as a function of the spatial separation and the orientation of the pairs.
Several flavours of perturbation theory have been used to model these
basic ingredients
obtaining satisfactory agreement with $N$-body simulations at least for certain redshifts, tracers and scales \citep{reid11,reid+12,Carlson+13, wang14,white14,vlah16,kopp16}. 

Analytical considerations and numerical simulations show that 
the los pairwise velocity distribution is strongly non-Gaussian 
at all scales, being characterized by a net skewness and approximately exponential tails \citep{efstathiou+88,Fisher+94,juszkiewicz+98, sc04}. 
Different strategies have been employed to explain this shape
ranging from the halo model \citep{sheth96, sheth01, ti07_2} 
to the superposition of environment-dependent Gaussian or quasi-Gaussian distributions 
\citep{bi1, bi2}.
Given the non-Gaussian nature of the velocity PDF,
the GSM corresponds to a cumulant expansion truncated at second order \citep{reid11}. 
An approximate extension to the third cumulant has been presented using the Edgeworth expansion around a Gaussian at first order \citep{uh15}. 

After discussing
the limitations of the GSM,
we introduce a new parameterization for the pairwise velocity PDF and show that
it accurately reproduces the ouptut of $N$-body simulations both at the level of particles
and haloes as well as 
for their 2-point correlation functions in redshift space.
There is a growing interest in extending RSD studies to smaller scales as a test of
modified gravity and interacting dark energy models \citep[e.g.][]{Jennings12,marulli12, hellwing+14,Taruya+14,Zu+14,Xu15,Barreira+16,Sabiu+16,Arnalte-Mur+17}.
These future developments provide the main motivation for our work. 

The paper is organized as follows. 
In Sections~\ref{sec:sim} and \ref{sec:rsd}, we introduce the suite of $N$-body  simulations used for our study and the basic principles of redshift-space distortions respectively.
In Section~\ref{sec:str}, we present an original derivation of the streaming model starting from the 2-particle distribution function in phase space. We show that
the model is regulated by different equations depending
on whether one is considering ordered or unordered pairs. 
Here, we also review
the applications of the streaming
model to galaxy redshift surveys
and test the basic assumptions of the GSM against $N$-body simulations.
In Section~\ref{stats}, we illustrate how the relative los velocity and its cumulants are connected to the (isotropic) radial and tangential components of the pairwise velocity vector. We then characterise
the scale dependence of
the first 4 cumulants for
dark-matter (DM) particles and haloes
extracted from our numerical simulations. This allows us to discuss
the pairwise velocity bias for the haloes.
Further, we use the $N$-body simulations to decompose
the matter pairwise velocity distribution into the contributions
generated by DM haloes of different masses.
In Section~\ref{sec:phe}, we introduce the generalized hyperbolic distribution to model the PDF of the los pairwise velocity and show that it vastly improves upon previous approximations for both
DM particles and haloes.
Finally, in Section~\ref{sec:summary}, we summarise our main achievements and conclude.
\begin{table*}
\caption{Cosmological parameters characterizing our $N$-body simulations.}
\begin{tabular}{cccccccc}
\hline
Name & $f_{\mathrm{NL}}$&$h$      & $\sigma_8$ & $n_{\mathrm{s}}$ & $\Omega_{\mathrm{m}}$ & $\Omega_{\mathrm{b}}$ & $\Omega_{\Lambda}$ \\
\hline
W0 & 0 &0.7010 & 0.8170   & 0.9600         & 0.2790              & 0.0462              & 0.7210   \\
W$f_{\mathrm{NL}}$ & $\pm$27,$\pm$80 &0.7010 & 0.8170   & 0.9600         & 0.2790              & 0.0462              & 0.7210   \\
P0 & 0 &0.6774 & 0.8159   & 0.9667         & 0.3089              & 0.0486              & 0.6911   \\
\hline
\end{tabular}
\label{table:cos}
\end{table*}

\section{$N$-Body simulations}\label{sec:sim}
Our study combines analytical and numerical work.
For the latter, we consider six $N$-body simulations run with 
the code \textsc{gadget-2} \citep{sp01,sp05}. 
As a reference, we use the zero-redshift output of the simulation that
was labelled 1.0 in \cite{pi10} and we now dub W0. This run
follows the formation of the large-scale structure from Gaussian initial conditions within a periodic cubic box with a side of 1200 $h^{-1} \mathrm{Mpc}$. It assumes a flat $\Lambda$CDM cosmology with the best-fitting parameters determined by the 5-yr analysis of the WMAP mission \citep{ko09} and
considers 1024$^3$ particles with a mass of $1.246 \times 10^{11} h^{-1} \mathrm{M}_\odot$
(see Pillepich et al. 2010 for further details).
To study the redshift evolution of our results, we
consider other snapshots extracted from the same run.
Furthermore, in order to explore the sensitivity of our findings to the
underlying cosmological model and to the properties of the linear density perturbations, we use other five simulations.
Four of them (W-80, W-27, W+27, W+80) are also presented in \citet{pi10}. Their only difference
with respect to our reference run is the presence of non-Gaussian
initial conditions of the local type with $f_{\mathrm{NL}}=-80, -27, +27$ and $+80$. 
The remaining simulation (P0), instead, has been run specifically
for this work and assumes the best-fitting cosmological parameters from the Planck mission \citep{Planck15I}. 
In all cases, we use 
the same box and softening lengths as well as the same number of particles and initial redshift. This implies that the P0 run has
a slightly higher particle mass, $1.379 \times 10^{11} h^{-1} \mathrm{M}_\odot$.
A comparison of the cosmological parameters used in the simulations is given in Table.~\ref{table:cos}.

We identify DM haloes using the {\sc{rockstar}} 
halo finder \citep{be13} in the default configuration  and only consider bound objects containing at least 100 particles within the virial radius.

\section{Redshift-space distortions}
\label{sec:rsd}
\begin{figure*}
		\centering
		\includegraphics[scale=0.8]{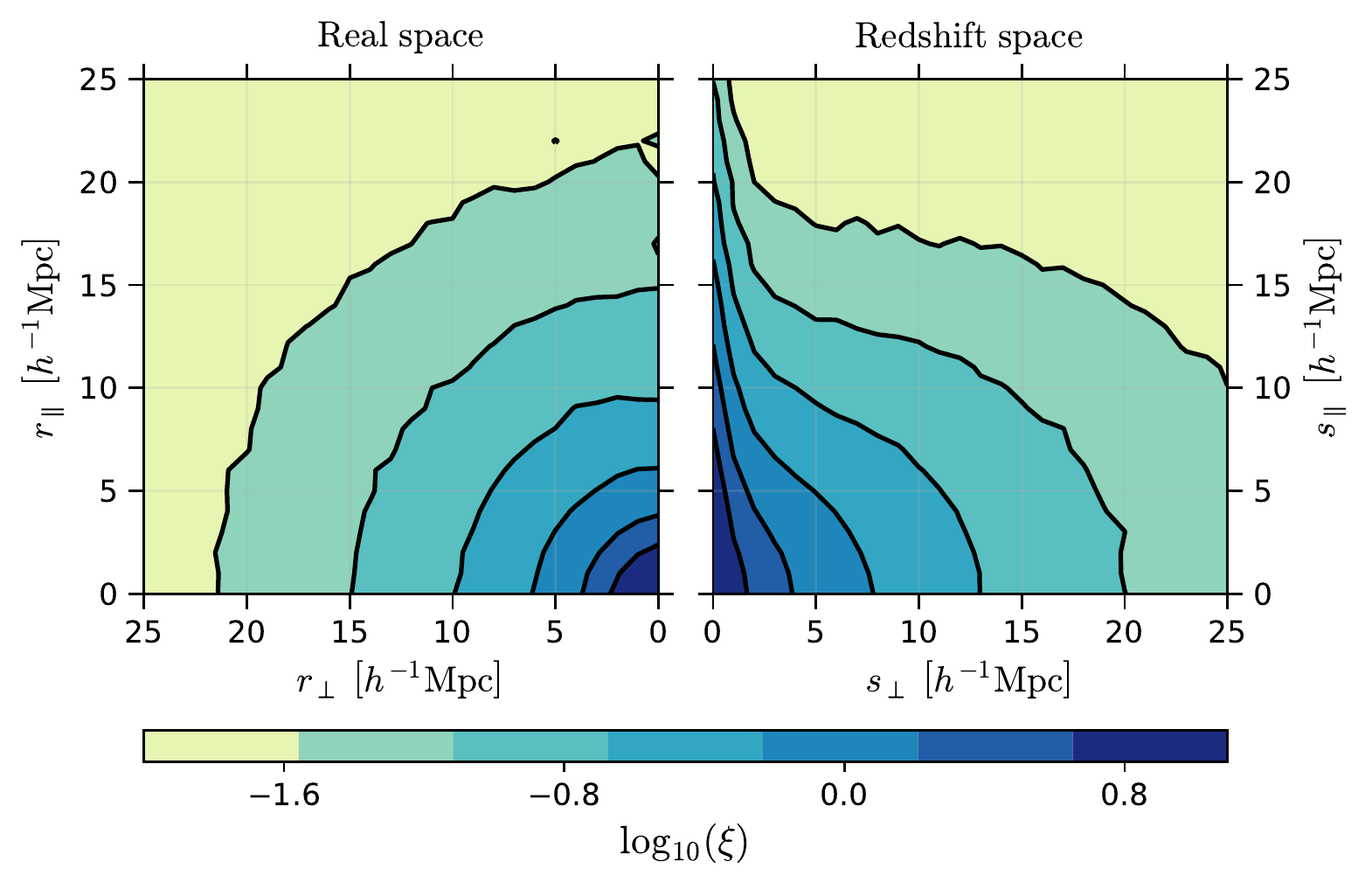}
 		\caption{Two-point correlation function in real (left) and redshift space (right) for the DM particles in the W0 simulation at $z=0$. The contour levels correspond to values of $\xi$ and $\xi_{\rm s}$ ranging from 0.08 (outermost) to 2.56 (innermost) and differing by a factor of 2 between two consecutive levels. To facilitate comparisons, the same levels are used in all figures showing correlation functions.}
		\label{fig:mi_corr}
	\end{figure*}

We generally use redshift as a distance indicator assuming a homogeneous model of cosmic expansion with instantaneous scale factor $a$ and Hubble parameter $H$.
However, in the presence of peculiar velocities, 
we need to distinguish between the redshift-inferred distance of a generic
tracer of the large-scale structure of the Universe
(e.g. a galaxy or a galaxy cluster) 
$\boldsymbol{x}_{\rm s}$ and its true comoving distance $\boldsymbol{x}$.
In the distant-observer (or plane-parallel) approximation \citep{hamilton98},
\begin{equation}
	\boldsymbol{x}_{\rm s} = \boldsymbol{x} + (\boldsymbol{\varv}\cdot\hat{\boldsymbol{z}})\,\hat{\boldsymbol{z}} \; ,
    \label{zmap1}
	\end{equation}
where $\boldsymbol{\varv}$ denotes the peculiar velocity divided by the factor $aH$
and $\hat{\boldsymbol{z}}$ denotes the los direction.
The spurious displacement along the los distorts the clustering pattern of the tracers in redshift space from its actual configuration in real space. 
Let us consider two tracers with real-space separation $\boldsymbol{r}=\boldsymbol{x}_2-\boldsymbol{x}_1$. The los component of their separation in redshift space is then
\begin{equation}
		s_{\parallel} = (\boldsymbol{x}_{{\rm s}_2} - \boldsymbol{x}_{{\rm s}_1})\cdot \hat{\boldsymbol{z}}=r_{\parallel} + w_{\parallel} \; ,
        \label{zmap2a}
\end{equation}
where $r_{\parallel} = \boldsymbol{r}\cdot\hat{\boldsymbol{z}}$  and $w_{\parallel} =(\boldsymbol{\varv}_{2} - \boldsymbol{\varv}_{1})\cdot\hat{\boldsymbol{z}}$,
while the transverse separation remains unchanged, i.e. 
\begin{equation}
\boldsymbol{s}_{\perp}=\boldsymbol{r}_{\perp}\;.
\label{zmap2b}
\end{equation}
If the spatial distribution of the tracers is statistically homogeneous in real space and their two-point correlation
function $\xi(r)$ is invariant under rotations of the separation vector $\boldsymbol{r}$, 
it follows that the correlation function in redshift space $\xi_{\mathrm{s}}(s_\perp,s_\parallel)$ is anisotropic between the parallel and transverse components
and does not depend on $s=(s_\parallel^2+s_\perp^2)^{1/2}$.
An illustrative example is shown in Fig.~\ref{fig:mi_corr} where we compare $\xi$ and $\xi_{\mathrm{s}}$ for the DM
particles in our $N$-body simulation (we have used one of the box axes as the los direction).
Note that the iso-correlation contours for $\xi_{\mathrm{s}}$ are elongated along $s_\parallel$
at small transverse separations as a manifestation of the FoG effect and are squashed 
towards the bottom at large $s_\perp$ because of coherent infall motions as described in \citet{kaiser87}.
\section{The streaming model}\label{sec:str}
The streaming model has been introduced to map $\xi(r)$ on to $\xi_{\mathrm{s}}(\boldsymbol{s})$
\citep{pe80,fi95, sc04}.
Its basic equation is generally written as
\begin{equation}
1+ \xi_\mathrm{s}(\boldsymbol{s}_{\perp},s_{\parallel}) = 
\int_{-\infty}^{+\infty}\left[ 1 + \xi\left(r\right) \right]\,
		\mathcal{P}_{w_{\parallel}}(s_{\parallel} - r_{\parallel} \mid \boldsymbol{r})\, \dif r_{\parallel} \;,
        \label{streamingclassic}
\end{equation}
where $r^2=s_\perp^2+r_\parallel^2$ and $\mathcal{P}_{w_{\parallel}}$ denotes the distribution function of the
pairwise los velocity at fixed real-space separation vector $\boldsymbol{r}=(\boldsymbol{s}_\perp,r_\parallel)$
(i.e. a PDF which is differential in $w_\parallel$).
In the remainder of this section,
we will show that this classic result
is exact provided that ordered pairs are used to define the correlation function and
the correct definition of the velocity PDF is employed. 

\subsection{Ordered and unordered pairs}
Two-point correlation functions are statistics of tracer pairs.
A pair is a set composed of two elements. Still, we can define two kinds of pairs.
If the order in which the elements appear in the pair is important (i.e. it makes
sense to define a first element and a second element),
we speak of an ordered pair (or 2-tuple).
On the other hand, if the order does not matter, we speak of an unordered pair (or pair set). 
Ordered pairs can be represented by directed graphs (and vice versa), e.g. $A\rightarrow B \neq B\rightarrow A$, and unordered pairs by undirected graphs, e.g. $A$ --- $B= B$ --- $A$.
From a set of $N$ discrete objects, we can form $N\,(N-1)$ ordered pairs and $N\,(N-1)/2$ unordered pairs.

If the two-point correlation function is built out of ordered pairs, the spatial separations
$r_\parallel$ and $s_\parallel$ must be signed numbers (while $s_\perp=r_\perp\geq 0$
as they give the magnitude of the two-dimensional vector $\boldsymbol{s}_\perp=\boldsymbol{r}_\perp$).
On the other hand, for unordered pairs, $r_\parallel$ and $s_\parallel$ are unsigned numbers.

\begin{figure*}
		\centering
		\includegraphics[scale=0.75]{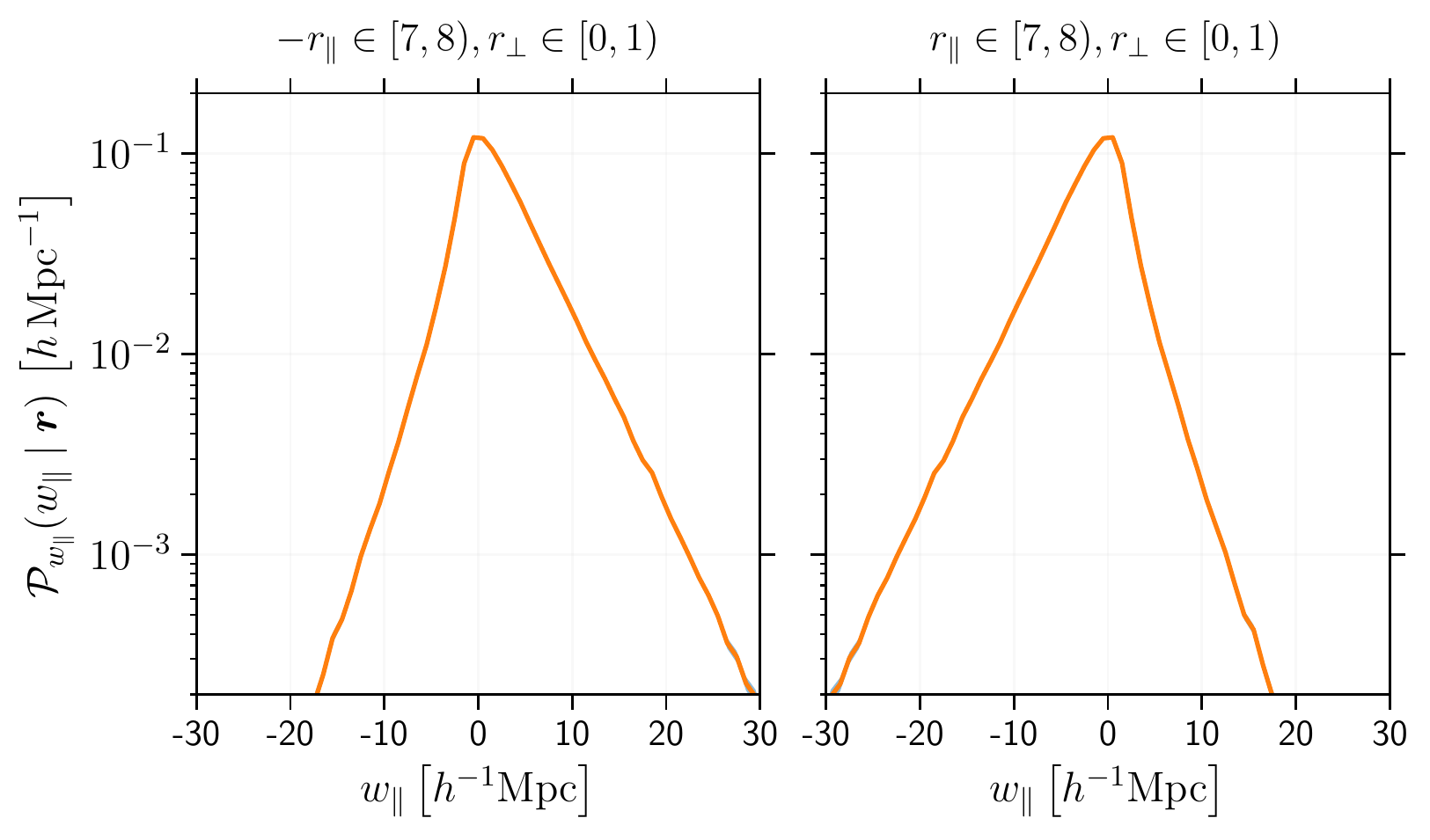}
		\caption{PDF of the relative los velocity for ordered pairs with fixed real-space separation (as indicated by the top labels in units of $h^{-1} \mathrm{Mpc}$).  The left- and right-hand side panels correspond to a particle exchange and show the symmetry $\mathcal{P}_{w_{\parallel}}(w_{\parallel}\mid r_{\perp},r_{\parallel}) = \mathcal{P}_{w_{\parallel}}(-w_{\parallel}\mid r_{\perp},-r_{\parallel})$.}
		\label{fig:mi_wp}
        \end{figure*}	
        
\subsection{The streaming model for ordered pairs}
\label{sec:order}    

Let us consider a system of $N$ particles with instantaneous comoving positions $\bx_i$ and rescaled peculiar velocities $\bv_i$ (where the subscript $i$ identifies the
different particles). Following a standard procedure in classical statistical mechanics, we introduce the one- and two-particle phase-space densities \citep{Yvon35}
\begin{align}
f_1(\boldsymbol{x},\boldsymbol{\varv})=\langle \hat{f}_1(\boldsymbol{x},\boldsymbol{\varv})\rangle&=\langle\sum_{i=1}^N  \delta_{\rm D}^{(3)}(\boldsymbol{x}-\boldsymbol{x}_i)
\,\delta_{\rm D}^{(3)}(\boldsymbol{\varv}-\boldsymbol{\varv}_i)\rangle \;,\label{f1def}\\
f_2(\boldsymbol{x}_{\rm A},\boldsymbol{x}_{\rm B},\boldsymbol{\varv}_{\rm A},\boldsymbol{\varv}_{\rm B})&=\langle \sum_{i=1}^{N}\sum_{j\neq i}
\delta_{\rm D}^{(3)}( \boldsymbol{x}_{\rm A}-\boldsymbol{x}_i)\,
\delta_{\rm D}^{(3)}(\boldsymbol{x}_{\rm B}-\boldsymbol{x}_j)\nonumber\\
&\,\,\,\,\,\,\,\,\,\delta_{\rm D}^{(3)}(\boldsymbol{\varv}_{\rm A}-\boldsymbol{\varv}_i)\,\delta_{\rm D}^{(3)}(\boldsymbol{\varv}_{\rm B}-\boldsymbol{\varv}_j)\rangle\;,
\end{align}
where $\hat{f}_1(\boldsymbol{x},\boldsymbol{\varv})$ is the discrete one-particle phase density (also known as the Klimontovich or the microscopic density), the brackets denote averaging over an ensemble of realisations and $\delta_{\rm D}^{(n)}(\bx)$
is the Dirac delta distribution in $\mathbb{R}^n$.
Note that $f_1$ is normalised to the total number of particles, i.e. $\int 
f_1\,\dif^3 x\,\dif^3\varv=N$ and
$f_2$ to the total number of ordered pairs, i.e. $\int f_2\, \dif^3x_{\rm A}\,\dif^3x_{\rm B}\,\dif^3\varv_{\rm A}\,\dif^3\varv_{\rm B}=
N\,(N-1)$.
By definition, the spatial two-point correlation function of the particles is
\begin{equation}
1+\xi(\bx_{\rm A},\bx_{\rm B})=\frac{\int f_2\,\dif^3 \varv_{\rm A}\, \dif^3 \varv_{\rm B}}{\left(\int f_1\,\dif^3 \varv\right)^2}\;.
\end{equation}
Assuming statistical isotropy (i.e. the invariance under rotations of the expectations
over the ensemble) implies that $f_1$ can only depend on $x^2, \varv^2$ and $\bx\cdot\bv$.
Requiring that ensemble averages are also invariant under translations (statistical
homogeneity) fixes the dependence of the one-particle distribution function to $f_1=\bar{n}\,F(\varv^2)$ where the constant
$\bar{n}=\int f_1\,\dif^3\varv$
gives the mean particle number density per unit volume and $F$ is an arbitrary function
such that $4\pi\int F(\varv^2)\, \varv^2\, \dif\varv=1$.
Under the same assumptions and for $N\gg 1$, 
$\int f_2\, \dif^3\varv_{\rm A}\,\dif^3\varv_{\rm B}=\bar{n}^2\,[1+\xi(r)]$. 

Our goal is to introduce a new set of distribution functions which are defined in
`redshift phase space'. This can be easily achieved by performing the change of
coordinates given in equations (\ref{zmap1}), (\ref{zmap2a}) and (\ref{zmap2b}):
\begin{align}
g_1(\bx_{\rm s},\bv)&= f_1(\bx_\perp,x_{{\rm s}\parallel} -\varv_\parallel,\bv)=\bar{n}\,F(\varv^2)\;,\\
g_{2}(\boldsymbol{s}_{\perp},s_{\parallel},\boldsymbol{\varv}_{\rm A},\boldsymbol{\varv}_{\rm B}) &= f_2(\boldsymbol{s}_\perp,s_\parallel-\varv_{{\rm B}\parallel}+\varv_{{\rm A}\parallel}, \boldsymbol{\varv}_{\rm A},\boldsymbol{\varv}_{\rm B})\;. 
\label{g2tof2}
\end{align}
The spatial two-point correlation function in redshift-space is then
\begin{equation}
1+\xi_{\mathrm s}(\boldsymbol{s}_\perp,s_\parallel)=\frac{\int g_2\,\dif^3 \varv_{\rm A}\, \dif^3 \varv_{\rm B}}{\left(\int g_1\,\dif^3 \varv\right)^2}=
\frac{\int g_2\,\dif^3 \varv_{\rm A}\, \dif^3 \varv_{\rm B}}{\bar{n}^2}
\;.
\label{xis-g2}
\end{equation}
We now rewrite the rhs of equation (\ref{g2tof2}) as
\begin{equation}
\int_{-\infty}^{+\infty}
f_2(\boldsymbol{s}_\perp,s_\parallel-w_\parallel, \boldsymbol{\varv}_{\rm A},\boldsymbol{\varv}_{\rm B})\,\delta^{(1)}_{\rm D}(w_\parallel-
\varv_{{\rm B}\parallel}+\varv_{{\rm A}\parallel})\,\dif w_\parallel\;,
\end{equation}
and we substitute it in equation (\ref{xis-g2}). After multiplying the rhs of the resulting
equation by the ratio
\begin{equation}
\frac{\bar{n}^2\,\left[1+\xi\left(\sqrt{s_\perp^2+(s_\parallel-w_\parallel)^2}\right)\right]}{\displaystyle \int f_2(\mathbf{s}_\perp,s_\parallel-w_\parallel,\boldsymbol\varv_{\rm A},\boldsymbol\varv_{\rm B}) \dif^3 \varv_{\rm A}\,\dif^3 \varv_{\rm B}}\;,
\end{equation}
(which is identically equal to one),  
we re-arrange the terms to define 
the relative line-of-sight velocity PDF for ordered pairs with real-space separation $\boldsymbol{r}=(\boldsymbol{s}_\perp,s_\parallel-w_\parallel)$ as
\begin{eqnarray}
\mathcal{P}_{w_{\parallel}}\!\!\!\!\!\!\!&(&\!\!\!\!\!\!\!w_{\parallel}\mid \boldsymbol{r})=\frac{\int f_2\left(\boldsymbol{r},\bv_{\rm A},\bv_{\rm B}\right) \, \delta^{(1)}_{\mathrm{D}}(w_{\parallel}-\varv_{{\rm B}\parallel}+\varv_{{\rm A}\parallel})\, \dif^3 \varv_{\rm A} \,\dif^3 \varv_{\rm B}}{\int f_2\left(\boldsymbol{r},\bv_{\rm A},\bv_{\rm B}\right) \dif^3 \varv_{\rm A} \dif^3 \varv_{\rm B}} \nonumber\\
&=&\!\!\!\!\!\frac{\int f_2\left(\boldsymbol{r},\bv_{\rm A},\bv_{\rm B}\right) \, \delta^{(1)}_{\mathrm{D}}(w_{\parallel}-\varv_{{\rm B}\parallel}+\varv_{{\rm A}\parallel})\, \dif^3 \varv_{\rm A} \,\dif^3 \varv_{\rm B}}{\bar{n}^2\,[1+\xi(r)]} \;,
\label{Pwdef}
\end{eqnarray}
and finally obtain
\begin{align}
1+ \xi_\mathrm{s}(\boldsymbol{s}_{\perp},s_{\parallel}) & =
\int_{-\infty}^{+\infty}  \left[ 1 + \xi\left(r\right) \right]\,
		\mathcal{P}_{w_{\parallel}}(w_{\parallel} \mid \boldsymbol{r}) \,\dif w_{\parallel} \nonumber\\
		& = \int_{-\infty}^{+\infty}  \left[ 1 + \xi\left(r\right) \right]
		\mathcal{P}_{w_{\parallel}}(s_{\parallel} - r_{\parallel} \mid \boldsymbol{r}) \dif r_{\parallel} \, \label{eq:ex_str} ,
\end{align}	
which coincides with equation (\ref{streamingclassic}). 
The moment-generating function of the random variable $w_\parallel$ is 
\begin{equation}
M_{w_\parallel}(t)=
\frac{\int e^{t(\varv_{{\rm B}\parallel}-\varv_{{\rm A}\parallel})}\,f_2\left(\boldsymbol{r},\bv_{\rm A},\bv_{\rm B}\right) \, \, \dif^3 \varv_{\rm A} \,\dif^3 \varv_{\rm B}}{\bar{n}^2\,[1+\xi(r)]}\;.
\label{mgf}
\end{equation}

Let us recap what we have done and achieved so far. Following a particle-based description and making use of the reduced distribution functions in phase-space,
we have demonstrated that the streaming model is exact in the distant-observer approximation
(and under the assumption of statistical homogeneity and isotropy in real space)
provided that equation (\ref{Pwdef}) is used to define $\mathcal{P}_{w_{\parallel}}$ or, equivalently, equation (\ref{mgf}) is used to calculate the moment-generating function.
It is worth stressing that our particle-based approach is completely rigorous also in multi-stream regions
(where particles with different velocities are present at the same spatial location
$\boldsymbol{x}$)
and fully accounts for density-velocity correlations.

Alternative derivations of the streaming equation have been presented by other authors adopting more restrictive assumptions. 
They are generally based on a macroscopic description 
obtained by coarse graining $\hat{f}_1(\boldsymbol{x},\boldsymbol{\varv})$ either in space (over patches of intermediate size between
the typical inter-particle separation and the cosmological scales of interest) or in time so that to erase discreteness effects and deal with a smooth
function. Formally, a perfectly smooth distribution function, $\bar{f}_1(\boldsymbol{x},\boldsymbol{\varv})$, is obtained
by taking the Rostoker-Rosenbluth fluid limit \citep{RR60},
i.e. by simultaneously letting $N\to \infty$ and the particle mass $m\to 0$ so that $Nm=$ constant.
This provides an approximated description of the system which is accurate until discreteness effects (particle collisions and correlations)
cannot be neglected any longer.
In fact, the microscopic, $n$-particle, and macroscopic
densities are distinct quantities that evolve very differently: 
$\hat{f}_1$ is exact and satisfies the Klimontovic equation, the $n$-particle densities, $f_n$, (which are exact ensemble averages) are solutions of the BBGKY hierarchy, while $\bar{f}_1$ is an approximation and fulfills the Vlasov equation.
Equipped with these definitions,
we are now ready to compare our results with previous derivations of the streaming equation.
\citet{sc04} used the language of statistical
field theory and characterized the matter content of the universe (in the single-stream regime) in terms of two continuous fields describing the density contrast $\delta(\bx)$ and the peculiar velocity $\boldsymbol{u}(\bx)$.
In our notation, this is equivalent to assuming that
$\bar{f}_1(\bx,\bv)=
\bar{n}\,[1+\delta(\bx)]\,\delta_{\rm D}^{(3)}[\bv-\boldsymbol{u}(\bx)]$.
Correlation functions were then computed by taking
ensemble averages of the product of fields evaluated  at different spatial locations.
This method has been generalised to multi-streaming by \citet{uh15} and \citet{Agrawal2017} 
 \citep[see also][]{seljak-mcdonald}, who wrote
$\bar{n}\,[1+\delta(\bx)] = \int \bar{f}_1(\bx,\bv)\,\dif^3 \varv$ before computing
the correlation function. 
In all cases, the authors were able to derive
equation (\ref{streamingclassic}).
However, since they rely on the collisionless fluid limit, all these approaches give a different
velocity PDF and moment-generating function than our equations (\ref{Pwdef}) and (\ref{mgf}). 
In fact, our
$f_2\left(\boldsymbol{r},\bv_{\rm A},\bv_{\rm B}\right)$ is replaced by the product
$\bar{f}_1(\boldsymbol{x}_{\rm A},\bv_{\rm A})\,\bar{f}_1(\boldsymbol{x}_{\rm B},\bv_{\rm B})$
which completely neglects velocity correlations.
While this is certainly an extremely good approximation for particle dark matter,
it does not hold true in general.
Note that our derivation is exact and
applies to any system
of particles (e.g. galaxies or their host haloes) without making assumptions regarding their interactions. 

\subsubsection{Symmetry under particle exchange}
By construction, the velocity PDF in equation (\ref{Pwdef}) is symmetric under particle exchange (A$\leftrightarrow$B)
or parity transformations, i.e. 
\begin{equation}
\mathcal{P}_{w_{\parallel}}(w_{\parallel}\mid \boldsymbol{r}) = \mathcal{P}_{w_{\parallel}}(-w_{\parallel}\mid -\boldsymbol{r})\;.
\end{equation}
In fact, the ordered pairs that can be formed with two particles 
equally contribute to $(w_\parallel, \boldsymbol{r})$ and $(-w_\parallel, -\boldsymbol{r})$.

In order to visualise the velocity PDF for the DM particles of our $N$-body simulation, 
we build an estimator for $\mathcal{P}_{w_\parallel}$ by replacing the ensemble average in the definition of $f_2$ with an average over the simulation box (accounting for periodic boundary conditions).
To speed the calculation up, we randomly sample $256^3$ particles and consider all
the ordered pairs between them. Each pair is classified in $1\times 1$ $(h^{-1}$ Mpc$)^2$ bins based on the values for $r_\parallel$ and $r_\perp$. Finally, we build an histogram
for $w_\parallel$ in each bin. An example is shown in Fig.~\ref{fig:mi_wp} where we examine real-space separations of $\pm r_\parallel \in [7,8)\,h^{-1}$ Mpc and $r_\perp\in[0,1)\, h^{-1}$ Mpc.
In each branch, the distribution clearly shows strongly asymmetric exponential tails. For 
$r_\parallel>0$, it has a negative skew meaning that the particles in the pairs tend to
approach each other. The mean and rms values are $-2.4\, h^{-1}$ Mpc and 
$5.2\, h^{-1}$ Mpc, respectively, while the mode is very close to zero.

	\subsection{The streaming model for unordered pairs}
	\label{sec:unordered}
   
    Before moving to the main goal of our work, we briefly discuss an alternative formulation of the streaming equation which highlights an interesting
    feature of RSD.

    $N$-body simulations provide valuable insights
    about the pairwise-velocity PDF. In this branch of the literature \citep{efstathiou+88,Zurek+1994,Fisher+94,sc04,ti07_1,reid11,zu13,bi1}, it is customary to use
the pairwise los velocity for unordered pairs,
\begin{equation}    
    w_{\rm los}=w_\parallel\,\frac{r_\parallel}{|r_\parallel|}=w_\parallel \,\mathrm{sgn}\left(r_{\parallel}\right)\;,
\end{equation}
whose sign
\color{black}
encodes physical information about the relative projected motion. If the elements of a pair approach (recede from) each other along the los, then $w_{\rm los}$ is negative (positive). 

\begin{figure}
		\centering
		\includegraphics[scale=0.7]{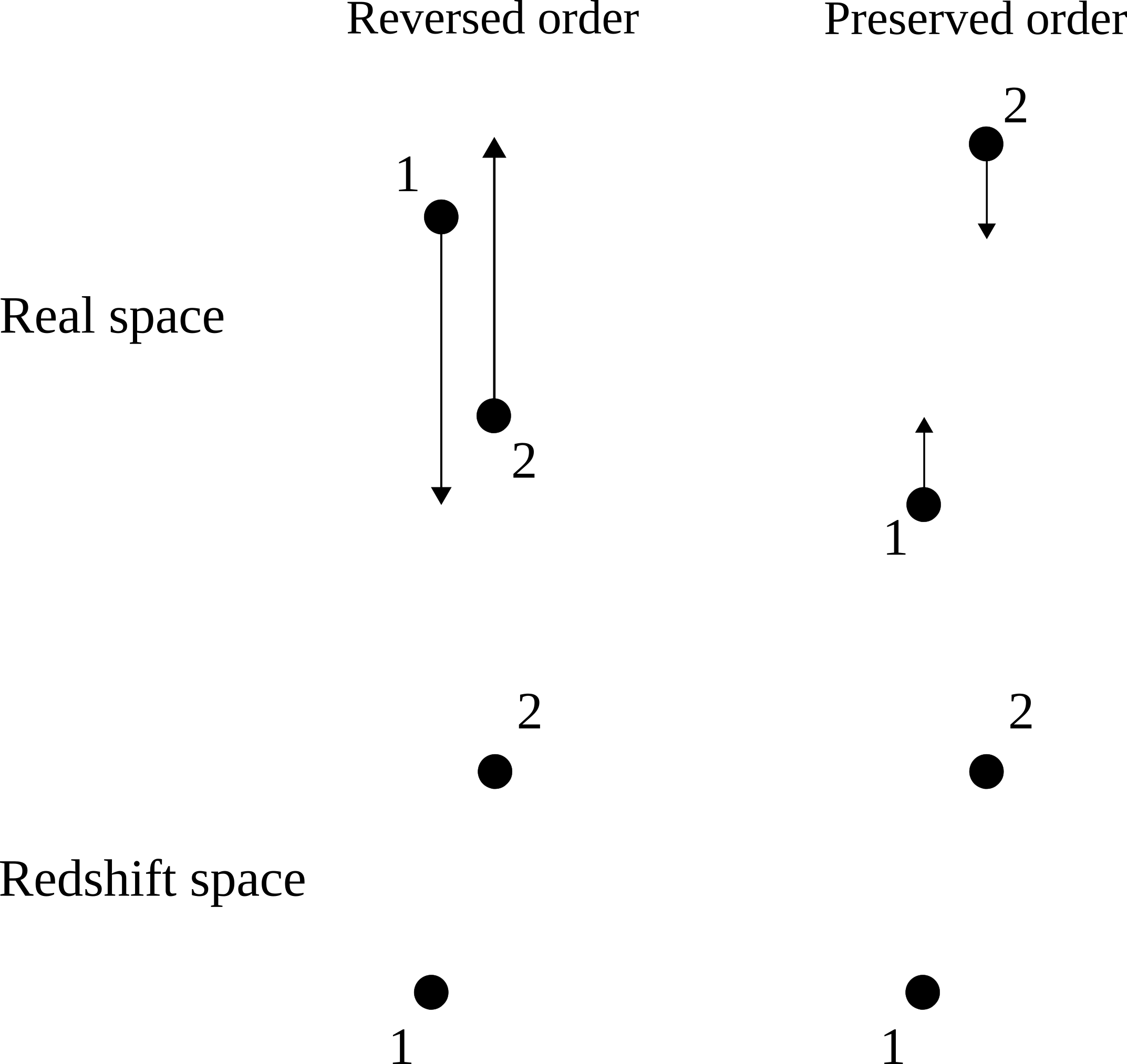}
		\caption{Schematic diagram illustrating how large relative infall velocities can reverse the order of a pair between real and redshift space (left-hand panel). The vertical arrows represent the los component of the individual velocities (assumed equal for simplicity). For comparison, the right-hand panel, shows a pair that preserves its order.}
		\label{fig:flips}
\end{figure}

It is worth stressing that the PDF of $w_{\rm los}$
can not be inserted into equation (\ref{streamingclassic}) without first making some changes. In this section, we clarify the issue and derive the basic equation of the streaming model for unordered pairs.
\color{black}
By construction, $w_{\rm los}$ does not depend on the order of the galaxy/particle pairs
and
\begin{equation}
\mathcal{P}_{w_{\rm los}}(w_{\rm los}|\mathbf{r})=\mathcal{P}_{w_{\rm los}}(w_{\rm los}|-\mathbf{r})=\mathcal{P}_{w_{\rm los}}(w_{\rm los}|r_\perp,|r_\parallel|)\;.
\end{equation}
For a generic $x$,
\begin{equation}
\mathcal{P}_{w_\parallel}(x\mid \boldsymbol{r})=
\begin{cases}
\mathcal{P}_{w_{\rm los}}(\phantom{-}x\mid r_\perp, |r_\parallel|)& \text{if}\  r_\parallel\geq 0\;,\\
\mathcal{P}_{w_{\rm los}}(-x\mid r_\perp, |r_\parallel|) & \text{if}\ r_\parallel<0\;,
\end{cases}
\end{equation}
i.e. $\mathcal{P}_{w_{\rm los}}$ coincides with the positive $r_\parallel$ branch
of $\mathcal{P}_{w_\parallel}$. Substituting in equation (\ref{streamingclassic}) and
breaking the integral into two parts running over the positive and negative values of $r_\parallel$ respectively, we obtain
\begin{equation}
1+ \xi_\mathrm{s}(s_{\perp},s_{\parallel}) =
\int_{0}^{+\infty}  \left[ 1 + \xi\left(r\right) \right]\,
		g(s_{\parallel}, |r_{\parallel}|, r_{\perp}) \,\dif |r_{\parallel}|\;,
\end{equation}	
with
\begin{equation}
g(s_{\parallel}, |r_{\parallel}|, r_{\perp})=\mathcal{P}_{w_{\rm los}}(-s_\parallel-|r_\parallel|| r_\perp,|r_\parallel|)+\mathcal{P}_{w_{\rm los}}(s_\parallel-|r_\parallel|| r_\perp,|r_\parallel|)\;.
\label{newdist}
\end{equation}
This is the correct equation of the streaming model for unordered pairs. It can be 
re-written in compact form using the signed $r_\parallel$,
\begin{equation} 
	1 + \xi_{\mathrm{s}}(s_{\perp},s_{\parallel}) = \int_{-\infty}^{+\infty}  \left[ 1 + \xi_{\mathrm{r}}(r) \right]
	\mathcal{P}_{w_{\mathrm{los}}}[(s_{\parallel} - r_{\parallel})\,\mathrm{sgn}(r_{\parallel}) \mid \mathbf{r}] \dif r_{\parallel} \; ,
    \label{eq:streaming}
	\end{equation}
which allows a direct comparison with equation (\ref{eq:ex_str}) for the ordered pairs.
As we illustrate in Fig. \ref{fig:flips},
the first term on the right-hand side of equation (\ref{newdist}) 
and the region with $r_\parallel<0$ in equation (\ref{eq:streaming})
refer to pairs that reverse their order between real and redshift space
(i.e. having separation $s_\parallel>0$ in redshift space and $-|r_\parallel|<0$ in real space).
On the other hand, the remaining terms are connected with pairs that preserve their ordering 
(i.e. both $s_\parallel$ and $r_\parallel$ are positive).

Pair reversal takes place more frequently at smaller real-space separations. The reason is twofold: 
i) smaller pairwise velocities are needed to swap the order of a pair when $r_\parallel$ is small; 
ii) the distribution of $w_{\rm los}$ shows a larger mean infall velocity and is 
more negatively skewed at small $r$.
A detailed discussion on the impact
of pair reversal on $\xi_s$ is presented 
in Appendix~\ref{app:revpairs}.

\subsection{Applications of the
streaming model}
\subsubsection{Pairwise velocities: mean infall and dispersion}
Most of the discussion on pairwise
velocities focus on their low-order statistical moments.
In our notation, the mean pairwise velocity is
defined as
\begin{equation}
\bv_{12}(\mathbf{r})=
\frac{\int 
(\bv_{\rm B}-\bv_{\rm A})\,
f_2\left(\boldsymbol{r},\bv_{\rm A},\bv_{\rm B}\right) \, \, \dif^3 \varv_{\rm A} \,\dif^3 \varv_{\rm B}}{\bar{n}^2\,[1+\xi(r)]}\;,
\end{equation}
which, in the literature, is often written in the single-stream fluid approximation as \citep[e.g.][]{juszkiewicz+98,reid11}
\begin{equation}
\bv_{12}(\mathbf{r})=\frac{\langle 
(\bv_{\rm B}-\bv_{\rm A})\,[1+\delta(\bx_{\rm A})]\,[1+\delta(\bx_{\rm B})]\rangle}
{1+\xi(r)}\;.
\end{equation}
By symmetry, $\bv_{12}(\mathbf{r})=\varv_{12}(r)\, \mathbf{r}/r$ and
the sign of $\varv_{12}(r)$ indicates whether
the elements of a pair, on average, approach each other or not.
For a set of self-gravitating particles,
$\varv_{12}(r)$ regulates the growth of the 2-point correlation function through
the pair-conservation equation that derives from
the BBGKY hierarchy \citep{DavisPeebles77}.

Similarly, 
the second moment of the pairwise velocities
defines a tensor of components,
\begin{equation}
\Sigma_{ij}(\boldsymbol{r})=
\frac{\int \Delta_i\,\Delta_j
f_2\left(\boldsymbol{r},\bv_{\rm A},\bv_{\rm B}\right) \, \, \dif^3 \varv_{\rm A} \,\dif^3 \varv_{\rm B}}{\bar{n}^2\,[1+\xi(r)]}\;,
\end{equation}
where $\Delta_i=(\bv_{\rm B}-\bv_{\rm A})\cdot \hat{\boldsymbol{x}}_i$ denotes the Cartesian
components of the pairwise velocity vectors with respect to a right-handed orthonormal basis $\hat{\boldsymbol{x}}_i$. 
By symmetry, $\Sigma_{ij}$
can be expressed in terms of a few scalar functions
\begin{equation}
\Sigma_{ij}(\boldsymbol{r})=\frac{\Bigl[\Xi(r)+\frac{2}{3}\sigma_{\varv}^2\Bigr]\,\delta_{ij}+\Bigl[\Pi(r)-\Xi(r)\Bigr]\,\frac{r_i\,r_j}{r^2}}{1+\xi(r)}\;,
\end{equation}
where $\sigma_{\varv}^2=\int \varv^2\,f_1\,\dif^3 \varv/\int f_1\,\dif^3 \varv$ denotes the mean square value of the peculiar particle velocity while 
$\Pi(r)$ and $\Xi(r)$ represent the contributions
of correlations for the velocity components
along $\hat{\boldsymbol{r}}$ and in the transverse directions, respectively
\citep{DavisPeebles77}.
The second moment of the pairwise-velocity component
along the separation vector is
\begin{equation}
\Sigma_{ij}(\boldsymbol{r})\,\hat{r}_i\,
\hat{r}_j=\frac{(2/3)\,\sigma_{\varv}^2+\Pi(r)}{1+\xi(r)}\;,
\end{equation}
while along any direction perpendicular to it
($\hat{\boldsymbol{t}}\cdot\hat{\boldsymbol{r}}=0$)
\begin{equation}
\Sigma_{ij}(\boldsymbol{r})\,\hat{t}_i\,
\hat{t}_j=\frac{(2/3)\,\sigma_{\varv}^2+\Xi(r)}{1+\xi(r)}\;.
\end{equation}
The second moment tensor is isotropic if
$\Pi(r)=\Xi(r)$. When this is not the case,
the velocity dispersion along the line of sight
depends on the pair orientation with respect to it
(see Section \ref{sec:rtl} for further details).

\subsubsection{Some historical remarks}
Since the advent of galaxy redshift surveys,
the streaming model has been a key tool to interpret
clustering data.
As we briefly mentioned in the introduction,
in its early applications, it was used to make inferences about galaxy motions
which are otherwise difficult to probe.
The anisotropies in $\xi_{\rm s}(s_\perp,s_\parallel)$ at small $s_\perp$ were
translated into a typical pairwise velocity
dispersion, $\sigma_{12}$, by assuming that
\citep{Peebles76,Peebles79}
\begin{equation}
\mathcal{P}_{w_{\parallel}}(w_{\parallel}\mid \boldsymbol{r}) = \frac{1}{\sqrt[]{2}\sigma_{12}}
\exp{\left(-\frac{\sqrt{2}\, |w_{\parallel}|}{\sigma_{12}}\right)}\;,
\label{expfit1}
\end{equation}
independently on the real-space separation. 
Quoting \citet{Peebles76}, the expression above
was `meant only as a simple fitting function with one free parameter'. It was actually motivated by
theoretical considerations and early $N$-body simulations showing that, on the spatial separations of interest: (i) the velocity PDF should be only weakly scale dependent, (ii) the second-moment tensor should be approximately isotropic and (iii) the pairwise velocity dispersion should be significantly larger than the mean. The small data sets available
at the time (containing a few hundred galaxy redshifts) were found to be consistent with these assumptions.

As samples grew bigger and the analysis was extended
to larger $s_\perp$, it was no longer possible
to neglect the so-called `streaming motions' i.e. the fact that the average relative velocity between galaxy pairs at fixed spatial separation does not vanish. 
If clustering is stable on small scales
(i.e. galaxy groups are in virial equilibrium), then $\varv_{12}(r)=-H\,r$. On the other
hand, for large separations, it is expected that $\varv_{12}(r)\to 0$. In between these asymptotic regimes, $\varv_{12}(r)<0$ and a net gravitational
infall should be observed. 
As a natural generalization of equation (\ref{expfit1}),
several authors then assumed that
\begin{equation}
\mathcal{P}_{w_{\parallel}}(w_{\parallel}\mid \boldsymbol{r})= \frac{1}{\sqrt[]{2}\sigma_{12}}
\exp{\left[-\frac{\sqrt{2}\, |w_{\parallel}-\bar{w}_\parallel(\boldsymbol{r})|}{\sigma_{12}}\right]}\;,
\label{eq:expfull}
\end{equation}
where $\bar{w}_\parallel(\boldsymbol{r})=\bv_{12}(\mathbf{r})\cdot \hat{\boldsymbol{z}}=\varv_{12}(r)\,\hat{\boldsymbol{r}}\cdot \hat{\boldsymbol{z}}
$ denotes
the line-of-sight component of the mean
pairwise velocity 
and $\varv_{12}(r)$ is written as a function of 
$\xi(r)$ using approximate solutions to the
pair-conservation equation calibrated against
$N$-body simulations 
\citep{DavisPeebles83,Bean83, Hale-Sutton+89,Mo+1993,Fisher+94,Zurek+1994, Marzke+1995, 
Loveday+1996, Shepherd+1997, Guzzo+1997, Jing+2002, Zehavi+2002, Li+2007}.

The larger volumes covered by the current
generation of redshift surveys lead to much more accurate measurements of galaxy clustering thus
providing strong motivation for better models.
Building upon the work by \citet{fi95},
\citet{reid11} proposed a Gaussian approximation for 
$\mathcal{P}_{w_{\parallel}}(w_{\parallel}\mid \boldsymbol{r})$
where the mean 
$\varv_{12}(r)\,\hat{\boldsymbol{r}}\cdot \hat{\boldsymbol{z}}$ and the scale-dependent dispersion $\sigma_{12}(\boldsymbol{r})$ are
computed using standard perturbation theory by assuming
that haloes are linearly biased tracers of the
matter. At the same time, the real-space correlation function of the galaxies
is evaluated using Lagrangian perturbation theory
and also including higher-order bias terms. Finally, in order to account for the incoherent motion of galaxies within their host haloes (and also to mitigate the imperfections of perturbative calculations), the actual variance of the Gaussian PDF is written as $\sigma^2(\boldsymbol{r})=\sigma_{12}^2(\boldsymbol{r})+\sigma^2_{\mathrm{FoG}}$
with $\sigma^2_{\mathrm{FoG}}$ a nuisance parameter over which one needs to marginalize.
The GSM has been later extended to use
velocity statistics for biased tracers computed within the Convolution Lagrangian Perturbation Theory \citep[CLPT,][]{wang14} and the Convolution Lagrangian Effective Field Theory
\citep[CLEFT,][]{vlah16}.
In its different versions, the GSM has been 
extensively applied to redshift surveys \citep{reid+12,Samushia+14,Alam+15,Chuang+17,Satpathy+17}.

\begin{figure}
	\centering
	\includegraphics[scale=0.57]{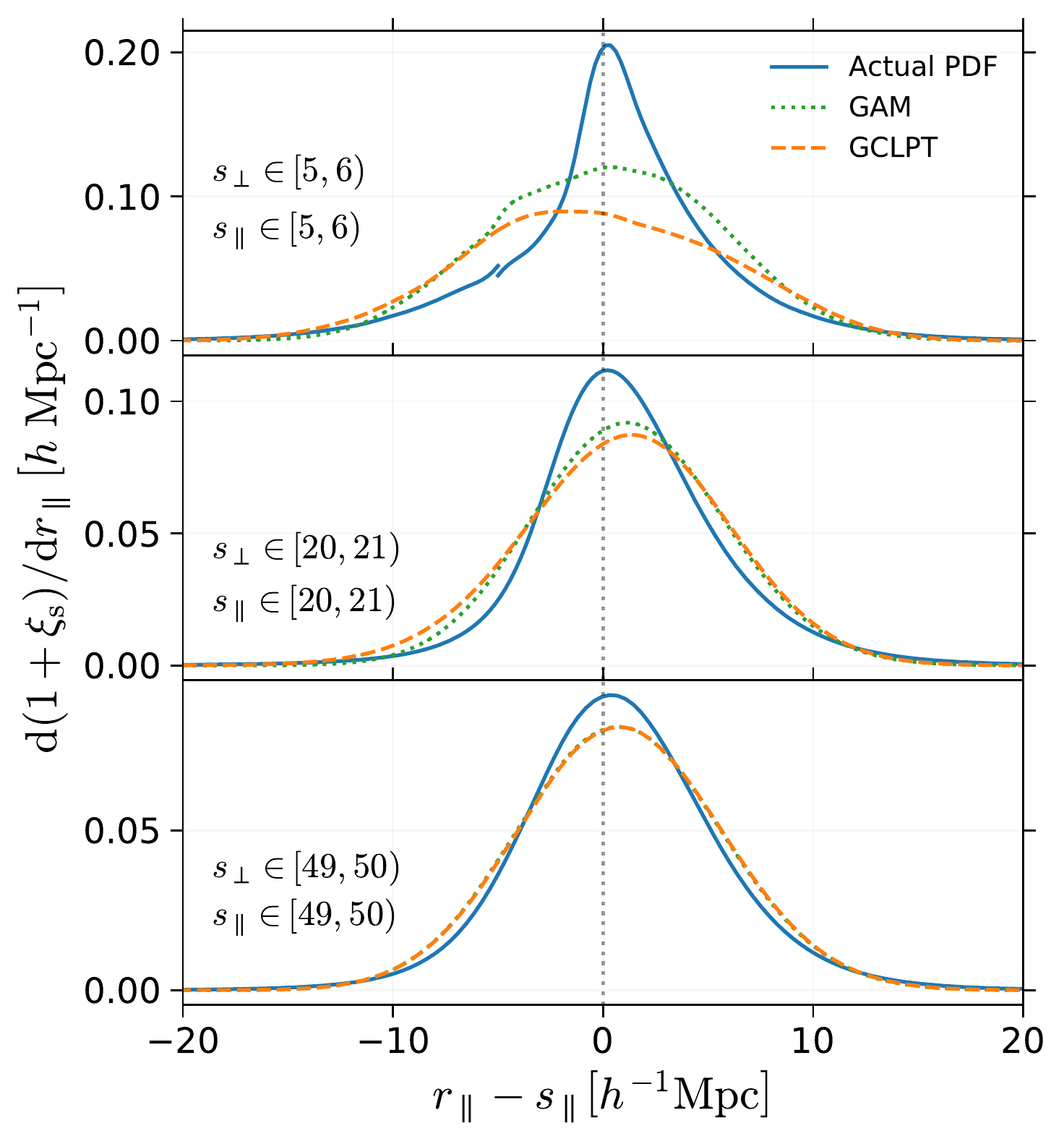}
	\caption{The integrand on the rhs of equation (\ref{streamingclassic}) is plotted
    as a function of $r_\parallel-s_\parallel$ for three
    redshift-space separation vectors (whose radial and transverse components are given in units of $h^{-1} \mathrm{Mpc}$). The bottom panel refers
    to scales that are usually treated perturbatively, the middle panel to intermediate scales, while the top panel considers separations in the fully non-perturbative regime. 
 In each case, we show three curves corresponding to different assumptions for the PDF of pairwise velocities. The solid curve corresponds to the actual data for the DM-particle pairs in the W0 simulation at $z=0$. The other curves are obtained by assuming a Gaussian PDF. For the dotted curve (GAM), the scale-dependent mean and variance are measured from the simulation
while, for the dashed curve (GCLPT), they are estimated using CLPT. In all cases, we use the same real-space correlation function, $\xi(r)$, which has been measured from the $N$-body simulation.
}
	\label{fig:clpt_xis}
\end{figure}
\subsubsection{Limitations of the Gaussian approximation}
The GSM
represents the state of the art in modelling RSD
at the 2-point level.
Considering DM haloes with masses between $10^{12}$ and $10^{13}\,h^{-1}$ M$_\odot$ at low redshift,
\citet{wang14} showed that the two-point
correlation $\xi_{\mathrm{s}}(\mathbf{s})$ predicted by the CLPT-based GSM agrees with $N$-body simulations to better than a few per cent for redshift-space separations larger than $\sim20\,h^{-1}$ Mpc
which are not too closely aligned with the line of sight \citep[similar results are obtained for the CLEFT-based GSM,][]{vlah16}.
The accuracy of the model degrades rapidly on smaller scales. We find similar trends when we 
study the correlation of DM particles, although, in
this case, the model already departs from the simulations on substantially larger scales (see below for more quantitative information).
It is instructive to investigate the origin of such
behaviour and clarify the implications of the Gaussian ansatz for the pairwise-velocity PDF.
To this goal, we focus on the DM
particles in the W0 simulation at $z=0$ as they
form a much larger sample than the DM haloes and are thus less affected by statistical noise.
In Fig.~\ref{fig:clpt_xis}, we plot the integrand
appearing in the rhs of equation (\ref{streamingclassic}) as a function of $r_\parallel-s_\parallel$ for three redshift-space
separation vectors and using three different
`models' for $\mathcal{P}_{w_{\parallel}}(w_{\parallel}\mid \boldsymbol{r})$.
The solid line is obtained by measuring the velocity
PDF at different pair separation vectors directly from the simulation. The dotted line (GAM) corresponds to assuming a Gaussian PDF with the actual scale-dependent mean and variance measured in the simulation. Finally, the dashed line (GCLPT) uses CLPT
to predict the mean and variance (up to an additive constant calibrated using simulations, see Section \ref{stats} and \citet{wang14} for further details) of the Gaussian PDF. 
In all cases,
we do not
compute $\xi(r)$ perturbatively but we measure
it directly from the simulation.
The top panel shows that both the Gaussian ansatz
for $\mathcal{P}_{w_{\parallel}}$ and the CLPT
calculations are very inaccurate at small scales. 
Integrating the different curves gives
$\xi_{\mathrm{s}}= 0.64$ for the $N$-body based PDF
while we obtain $\xi_{\mathrm{s}}= 0.70$ and $0.44$ for the GAM and GCLPT models, respectively.
The middle panel refers to a separation vector
with components $s_\parallel=s_\perp\simeq 20.5\, h^{-1}\textrm{Mpc}$. In this case, pairs with
$5\lesssim r_\parallel\lesssim 35\,h^{-1}\textrm{Mpc}$ contribute to $\xi_{\rm s}$
and those with $r_\parallel \simeq 20\,h^{-1}\textrm{Mpc}$ give the largest signal.
Both Gaussian approximations, however, reach their maximum for slightly larger values of $r_\parallel$ and give rise to less sharply peaked integrands
in the streaming equation.
Although the two curves based on the Gaussian approximations seem to be rather similar, once integrated, give rise to significantly different values of $\xi_{\rm s}$. We find
$\xi_{\mathrm s}=0.049$ when the actual moments are used and $\xi_{\mathrm s}=0.063$ for
the CLPT predictions. This difference
mainly originates from the regions on the lhs from the peak, i.e. for $r_\parallel<s_\parallel$ where perturbative calculations become less reliable.
Note that 
the actual value of the redshift-space correlation is $\xi_{\mathrm{s}}=0.048$ which nearly coincides
with the best Gaussian approximation in spite of the fact that the corresponding integrals in Fig.~\ref{fig:clpt_xis} appear to be quite different.
Even though we provided only one specific example here, we find that
this serendipitous coincidence holds true for a 
broad range of redshift-space separation vectors $\mathbf{s}$.
This result provides motivation for improving
the GSM by combining it with enhanced estimates
for the pairwise-velocity moments. However, the success of the GSM on these intermediate scales
appears to be fortuitous as the model does not
reproduce the correct shape of the integrand in the streaming equation.
The bottom panel shows that
the situation improves only slightly when we consider larger redshift-space separations. 
In this case, the CLPT calculations can be calibrated to accurately reproduce the first
two moments of the pairwise velocities and no obvious difference can be noted between the GAM and GCLPT approximations. However, even at a separation of
$s\simeq 70\,h^{-1} \mathrm{Mpc}$, the Gaussian
models for the velocity PDF cannot accurately reproduce the shape of the function $\mathrm{d}(1+\xi_{\mathrm{s}})/\mathrm{d}r_\parallel$ measured in the simulation. In spite of this, once the integral over $r_\parallel$ is performed, they give 
values for $\xi_{\mathrm s}$ which are accurate at the few per cent level.

The aim of this paper 
is to introduce a more general
fitting function for the pairwise-velocity PDF
that, once inserted in the streaming model, provides more accurate results
for the redshift-space correlation function than the
GSM.
The advantage is twofold: first, on large scales and
in the era of precision cosmology,
we would be able to make predictions that do not rely on fortuitous cancellations 
and, second, extending the accuracy of the streaming model to smaller spatial separations would allow us to probe modified gravity and interacting dark-energy models as mentioned in the introduction.
Related work has been presented by \citet{uh15}
and \citet{bi1, bi2} who accounted for skewness in the PDF by using the Edgeworth expansion around a Gaussian probability density 
and by superposing multiple Gaussian (or quasi-Gaussian) distributions, respectively.

	\begin{figure*}
	\centering
	\includegraphics[scale=0.55]{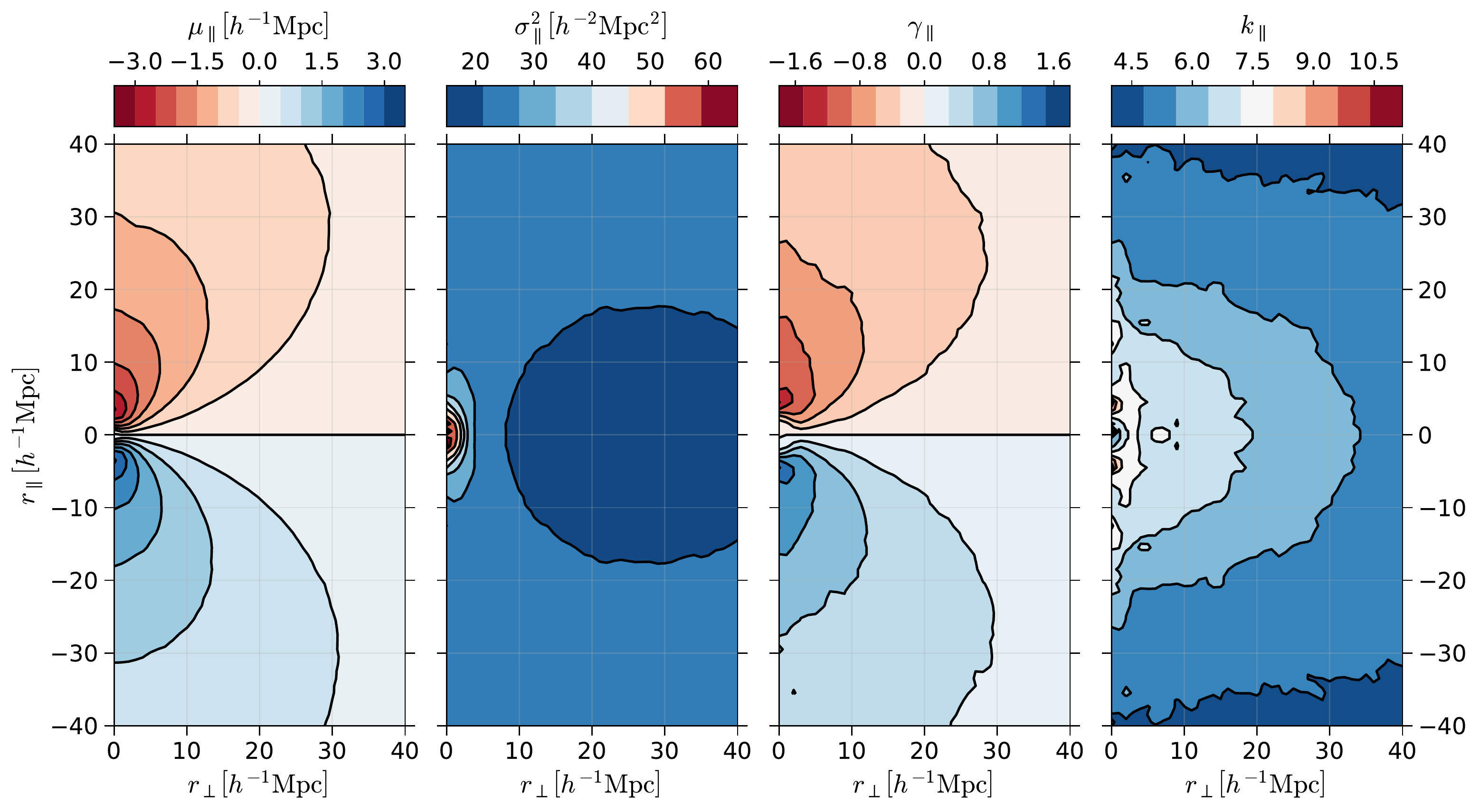}
	\caption{Normalised cumulants of the pairwise los velocity for ordered pairs of DM particles in the W0 simulation.}
	\label{fig:mo_wp}
\end{figure*}    
\section{Statistics of pairwise velocities}
The streaming model is formulated in terms of $w_{\parallel}$ or $w_{\mathrm{los}}$ while analytical calculations (perturbative or not) generally deal with the radial and transverse components of the pairwise velocities. In this section, we derive the relation between the cumulants of these different components. In order to provide some illustrative examples, we compute several velocity statistics for the DM particles
and the haloes in  the W0 simulation at redshift $z=0$.
\label{stats}    
 \subsection{Cumulants}    
In Fig.~\ref{fig:mo_wp}, we investigate how the
first four normalised cumulants of $\mathcal{P}_{w_\parallel}$
for the DM particles, 
depend on the real-space separation vector $\mathbf{r}$.   
Specifically, we consider
the mean $\mu_\parallel=\langle w_{\parallel}\rangle_{\rm c}$, the variance $\sigma^2_\parallel=\langle w_{\parallel}^2\rangle_{\rm c}
$, the skewness
$\gamma_\parallel=\langle w_{\parallel}^3\rangle_{\rm c}/\sigma_\parallel^3$ and the kurtosis $k_\parallel=\langle w_{\parallel}^4\rangle_{\rm c}/\sigma_\parallel^4$.
As expected, the odd cumulants undergo parity inversion as the sign of $r_{\parallel}$ changes while the even cumulants are parity invariant. 
Note that the cumulants of $\mathcal{P}_{w_{\rm los}}$ can also be read from Fig.~\ref{fig:mo_wp} by looking at the region with $r_\parallel>0$. In this case, the mean velocity and the skewness are always negative meaning that $\mathcal{P}_{w_{\rm los}}$ is asymmetric as it is more likely to find infalling pairs at the scales we consider
\citep[see also][]{sc04}.
The velocity PDF is leptokurtic (i.e. $k>3$) meaning that it has a sharper peak and heavier tails compared to a Gaussian distribution.
Although both $|\gamma|$ and $k$ decrease with increasing $r_\parallel$ and $r_\perp$,
$\mathcal{P}_{w_{\rm los}}$ always differs substantially from a Gaussian probability density.

 	\begin{figure*}
	\centering
	\includegraphics[scale=0.55]{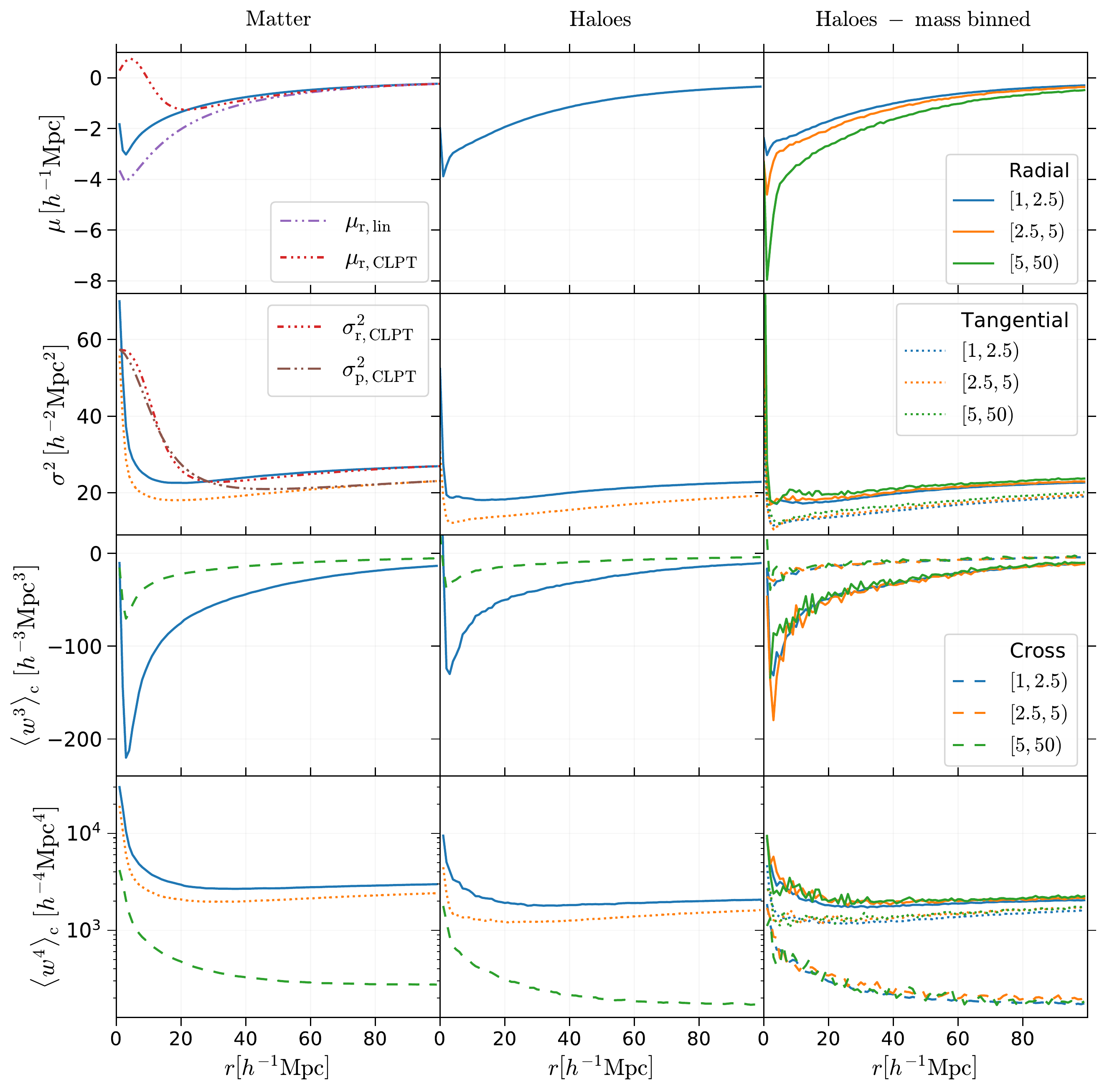}
	\caption{Cumulants and cross-cumulants of the radial and tangential pairwise velocities contributing to the first 4 normalised cumulants of $w_\parallel$, see equations (\ref{eq:v1}) - (\ref{eq:v4}). The solid, dotted and dashed lines represent the radial, tangential  and cross components, respectively. 
In the left column,  
all quantities have been measured from the velocities of DM particles in the W0 simulation at $z=0$. One-loop (and linear, for $\mu_{\rm r}$) CLPT predictions are also shown in the two topmost panels. A constant has been added to the CLPT results for $\sigma^2_{\rm r}$ and $\sigma^2_{\rm p}$ in order to match the simulation output at $r=100\,h^{-1}$ Mpc.
The middle column, shows the results
for the bulk velocities of the DM haloes identified in the same simulation snapshot. Finally,
the dependence of the results on the halo mass
is investigated in the right column using
different mass intervals
indicated in units of $10^{13}
h^{-1}$ M$_\odot$.}
	\label{fig:mo_radtan}
\end{figure*}

    \subsection{Radial, transverse and los pairwise velocities}
\label{sec:rtl}    
The peculiar shape of the contour levels in Fig.~\ref{fig:mo_wp} is mainly regulated by the angle that the pair separation forms with the los. This can be shown as follows.
The pairwise velocity of an ordered pair, $\boldsymbol{\varv}_2 - \boldsymbol{\varv}_1$, can be decomposed 
into radial (i.e. along the pair-separation vector) and transverse components:
\begin{align}	
	w_{\mathrm{r}} &= \left(\boldsymbol{\varv}_2 - \boldsymbol{\varv}_1\right)\cdot \hat{\boldsymbol{r}} \;, \\
    \mathbf{w}_{\mathrm{t}} &= \left(\boldsymbol{\varv}_2 - \boldsymbol{\varv}_1\right)-
    w_{\mathrm{r}}\,\hat{\mathbf{r}}\;.
    \end{align}
In a homogenous and isotropic universe, the statistical properties of $w_{\mathrm{r}}$
and $\mathbf{w}_\mathrm{t}$ only depend on $r$.
Introducing a preferential los direction, however, breaks the rotational invariance so that $\mathcal{P}_{w_\parallel}$ depends on both $r_\parallel$ and $r_\perp$. 
Let us consider the normal vector to the plane defined by the pair separation and the los,
\begin{equation}
		\boldsymbol{n} = \hat{\boldsymbol{r}} \times \hat{\boldsymbol{z}} \;,
\end{equation}	
and form a right-handed orthonormal basis using $\hat{\mathbf{r}}$, $\hat{\mathbf{n}}$ and the additional unit vector $\hat{\mathbf{e}}$. 
Only the component of $\mathbf{w}_{\mathrm{t}}$ perpendicular to $\boldsymbol{n}$ 
(and thus parallel to $\hat{\mathbf{e}}$),
\begin{equation}
	\boldsymbol{w}_{\mathrm{e}} = (\boldsymbol{w}_{\mathrm{t}}\cdot \hat{\boldsymbol{e}})\,\hat{\boldsymbol{e}}=    
    \left(\boldsymbol{\varv}_2 - \boldsymbol{\varv}_1\right) - w_{\mathrm{r}}\,\hat{\boldsymbol{r}} - (\mathbf{w}_\mathrm{t} \cdot \hat{\boldsymbol{n}})\,
\hat{\boldsymbol{n}}    \;, \end{equation}
contributes to $w_\parallel$.
By defining the angle $\theta$ so that $\cos\theta = \hat{\mathbf{r}}\cdot\hat{\boldsymbol{z}}=r_{\parallel}/r$, it follows that
$\hat{\mathbf{e}}\cdot\hat{\boldsymbol{z}}=\pm\sin\theta$ depending on the relative
orientation of the pair separation and the los. Eventually, we can write
	\begin{equation} \label{eq:los_vel_equivalence}
	w_{\parallel} = w_{\mathrm{r}} \cos\theta + w_{\mathrm{p}}\sin\theta \;,
	\end{equation}
 with
	\begin{equation}
		w_{\mathrm{p}} = \mathrm{sgn}(\hat{\boldsymbol{e}}\cdot \hat{\boldsymbol{z}})\,  w_{\mathrm{e}} \;.
	\end{equation}
The cumulants of $w_\parallel$ at fixed $r$ and $\theta$
can then be expressed
in terms of the cumulants and cross-cumulants of $w_{\mathrm{r}}$ and $w_{\rm p}$ at fixed $r$.
It follows from equation (\ref{eq:los_vel_equivalence}) that
	\begin{equation}
    \label{cumulants}
		\langle w_{\parallel}^n \rangle_{\rm c} = \sum_{i_1=1}^{2} \sum_{i_2=1}^{2} \dots\sum_{i_n=1}^{2}
        \langle w_{i_1} w_{i_2}\ldots w_{i_n} \rangle_{\rm c} \, ,
	\end{equation}
	where $w_1\equiv w_{\mathrm{r}} \cos \theta$ and $w_2\equiv w_{\mathrm{p}} \sin \theta$.
Terms involving odd powers of $w_{\mathrm{p}}$ vanish due to statistical isotropy
(i.e. the probability distribution of $w_{\mathrm{p}}$ is symmetric with respect to
zero, reflecting the fact that all orientations of $\hat{\boldsymbol{r}}$ with respect to $\hat{\boldsymbol{z}}$ are equally likely).
For the normalised cumulants shown in Fig.~\ref{fig:mo_wp}, we obtain:
\begin{align}\label{eq:v1}
		\mu_{\parallel} & = \mu_{\mathrm{r}}\, \cos\theta  \;, \\
        \label{eq:v2}
        \sigma^2_{\parallel} &= \sigma^2_{\mathrm{r}}\, \cos^2\theta + \sigma^2_{\mathrm{p}} \,\sin^2\theta \; ,\\
        \label{eq:v3}
\gamma_{\parallel} &=
 \sigma_\parallel^{-3}\,\left[\langle w^3_{\mathrm{r}}\rangle_{\rm c}\cos^2
\theta  + 3 \,\langle w_{\mathrm{r}}\,w^2_{\mathrm{p}}\rangle_{\rm c}\,\sin^2\theta \right]\cos\theta  \;,\\
\label{eq:v4}
      k_{\parallel} &= 
      \sigma_\parallel^{-4}\,\left\{\langle w^4_{\mathrm{r}}\rangle_{\rm c}\,\cos^4\theta  + \langle w^4_{\mathrm{p}}\rangle_{\rm c}\,\sin^4\theta +\right. \nonumber \\
		& \,\,\,\,\,\,\,\,6\left.\left[\langle w^2_{\mathrm{r}}\,w^2_{\mathrm{p}}\rangle_{\rm c} \right. 
		\left. - \,2\,\mu_{\mathrm{r}}\, \langle w_{\mathrm{r}}\,w^2_{\mathrm{p}}\rangle_{\rm c}\right]\cos^2\theta\sin^2\theta \right\}\;.
        \end{align}
\subsubsection{DM particles}
In the left column of Fig.~\ref{fig:mo_radtan}, we show the radial dependence of the different isotropic terms appearing in the equations above for the particles in the final snapshot ($z=0$) of the
W0 simulation.
In the top row, we compare $\mu_{\rm r}$ measured in our simulation with the predictions of linear
and one-loop Convolution Lagrangian Perturbation Theory \citep[CLPT][]{Carlson+13, wang14}.
In the second row, we plot $\sigma^2_{\mathrm{r}}$ and $\sigma^2_{\mathrm{p}}$
for the $N$-body particles and contrast them with the corresponding CLPT results at one loop (after shifting them vertically in order to match the simulation output at $r=100\,h^{-1}$ Mpc).
The figures indicate that state-of-the-art perturbative approaches qualitatively reproduce
the scale dependence of the lowest-order cumulants for $r>20-30\,h^{-1}$ Mpc. However,
they provide accurate predictions only on much larger scales.
Finally, the third and fourth rows present the various contributions to the skewness and kurtosis of $w_\parallel$. 
Note that the cross-cumulants of the correlated random variables $w_{\rm r}$ and $w_{\rm p}$ are sub-dominant but non-negligible.

\subsubsection{DM haloes and pairwise velocity bias}
\label{halobias}
\begin{figure}
\centering
\includegraphics[scale=0.57]{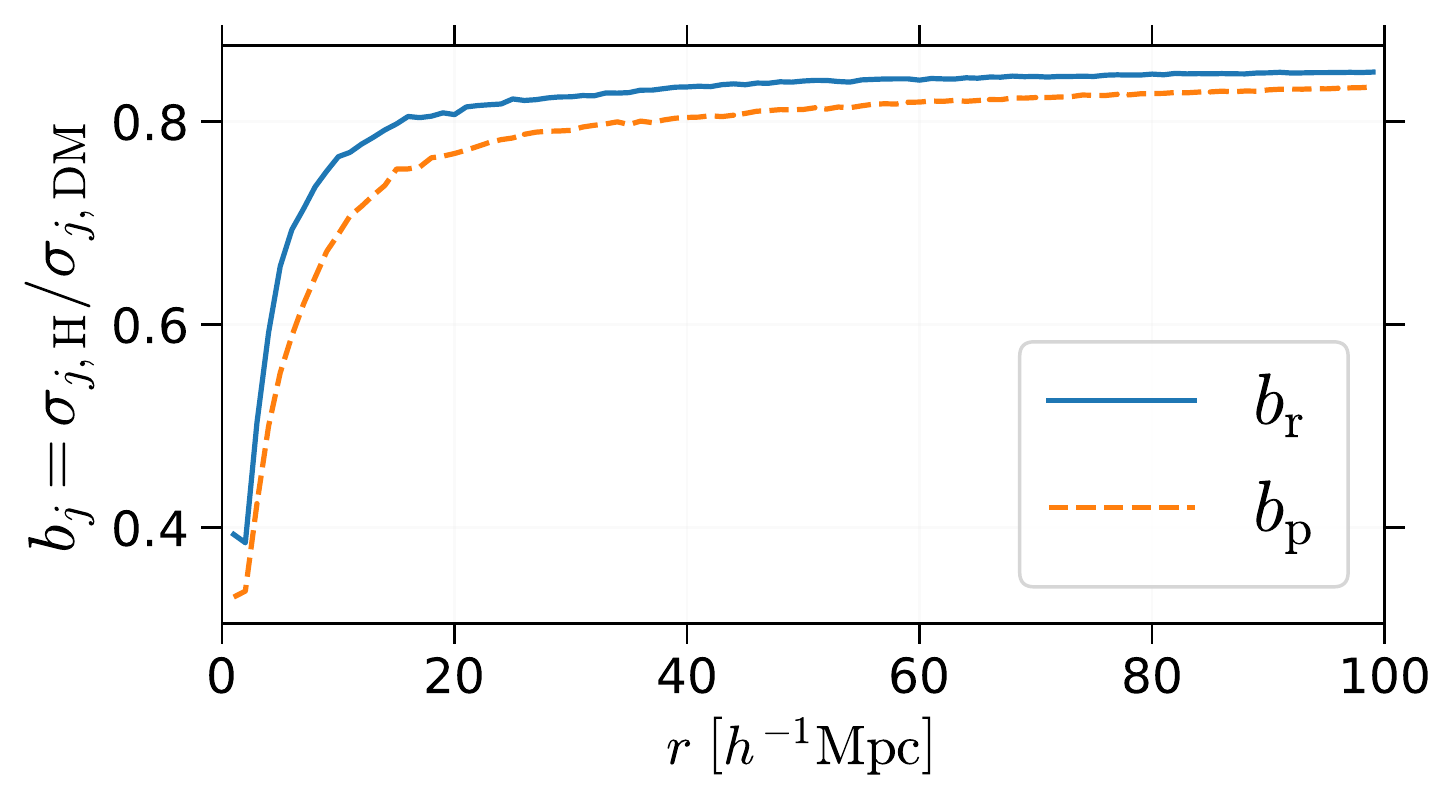}
	\caption{Pairwise-velocity bias for radial (solid) and tangential (dashed) motion.}
	\label{fig:pairwise_bias}
\end{figure}

We now contrast the previous results with the isotropic velocity statistics for the DM haloes identified at $z=0$ in the W0 simulation.
In the middle column of Fig.~\ref{fig:mo_radtan}, we show the cumulants and cross-cumulants for the radial and tangential pairwise velocities measured using the halo bulk velocities. 
It is evident that halo pairs (that trace peaks in the density field)
tend to approach
each other at a slightly greater velocity than
particle pairs (that also populate underdense regions where the velocity divergence is positive).
Haloes have also smaller velocity dispersions, i.e.
they trace a colder velocity field than the DM
\citep[see also][]{Carlberg+89,CouchmanCarlberg92,CenOstriker92,Gelb+94,Evrard+94,Summers+95,Colin+00}.
Following \citet{Carlberg94}, we introduce the
pairwise velocity bias as the ratio between
the halo and DM rms pairwise velocities. Its variation with the pair separation is shown in
Fig.~\ref{fig:pairwise_bias} for the radial
and the tangential components of the velocity.
In both cases, the `antibias' assumes values $\sim 0.8$ on large scales and 
becomes very strong for $r<15\,h^{-1}$ Mpc.
Fig.~\ref{fig:mo_radtan} shows that also
the higher-order cumulants for the haloes depart from those for the DM, especially at small separations.
This reflects the fact that the shape of the pairwise velocity PDF is different for DM particles and haloes.

In the right column of Fig.~\ref{fig:mo_radtan}, we investigate the
dependence of the velocity cumulants on halo mass.
More massive objects show larger mean infall velocities, dispersions, and fourth-order cumulants. On the other hand, we could not detect any mass dependence for the third-order cumulant. The latter,
however, is always negative indicating the
the halo velocity PDF is asymmetric (and thus non Gaussian) for all the separations and masses considered here. The amplitude of the fourth cumulant (invariably larger than 3)
also shows that the PDF is leptokurtic with heavy tails. 

\subsection{Dissecting the pairwise-velocity distribution}\label{sec:dis}
    
\begin{figure}
\centering
\includegraphics[scale=0.52]{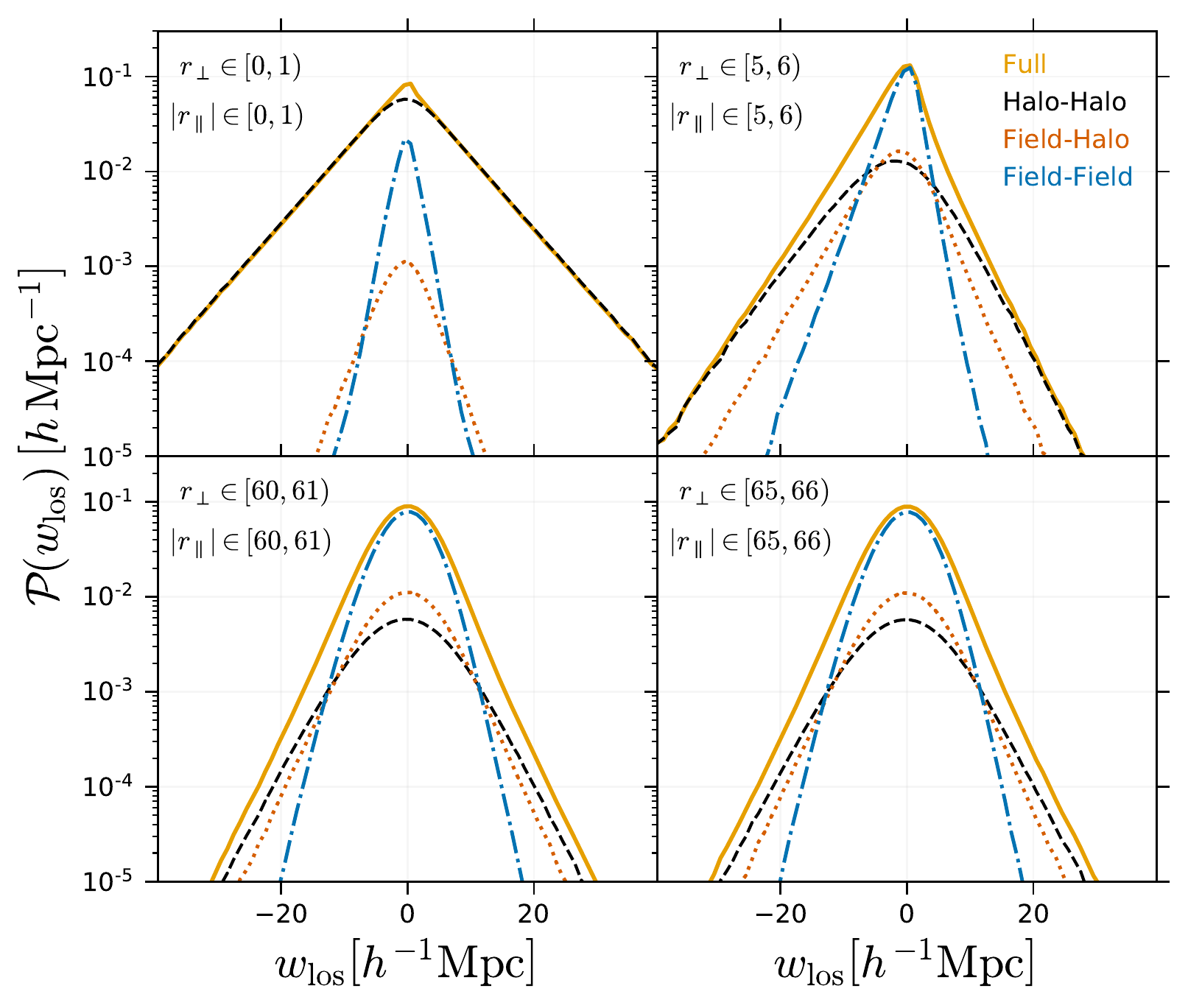}
\caption{The los pairwise-velocity distribution of the particles in the \textsc{W0} simulation at $z=0$ is
decomposed into simpler elements. 
The figure shows the contributions of halo-halo, field-halo and field-field pairs for six
different real-space separation vectors (expressed in $h^{-1} \mathrm{Mpc}$).}
\label{fig:rel_los_comp}
\end{figure}

The pairwise-velocity distribution 
of DM particles
is shaped by the highly non-linear physics of
gravitational instability and, for this reason, is very difficult to model analytically
starting from first principles.
A simplified approach relies on a phenomenological description that exploits the internal dynamics of DM haloes \citep[e.g][]{sheth96,sheth01,ti07_2}. 
To provide further insight into the importance of virialised structures, we investigate the halo contribution to $\mathcal{P}_{w_{\rm los}}$. 
We first classify the DM particles in our simulation according to whether they belong
to haloes or to the field. We then partition $\mathcal{P}_{w_{\rm los}}$ into the contributions of halo-halo, halo-field and field-field pairs. Note that the concept of `field
particle' is not absolute as it depends on the mass resolution of the simulation and the minimum halo mass which is considered. In the W0 simulation, we classify 69.3 per cent of the particles as belonging to the field at $z=0$. If we were resolving haloes with a mass
$M<1.2\times 10^{13}\,h^{-1} \mathrm{M}_\odot$, then the fraction of field particles would decrease. Basically, our `halo particles' account for the matter content of galaxy groups and clusters of galaxies.
We show a few examples in Fig.~\ref{fig:rel_los_comp}. 
For real-space separations that are smaller than the typical halo size, $r<1\,h^{-1}$ Mpc,
halo-halo pairs give the dominant contribution to $\mathcal{P}_{w_{\rm los}}(w_{\rm los}|{\mathbf{r}})$ for all $w_{\rm los}$ whereas the field-field pairs only matter at very low $w_{\rm los}$. 
For $r>1\,h^{-1}$ Mpc, instead, 
the wings of $\mathcal{P}_{w_{\rm los}}$ are regulated by the halo-halo pairs while the core of the PDF is determined by the field-field pairs. Halo-field pairs are always subdominant. 
The field-field term peaks at $w_{\rm los}\simeq 0$, is negatively skewed 
(although it becomes practically symmetric for $r_\parallel<1\,h^{-1}$ Mpc) and
shows rapidly decaying exponential wings.
Halo-halo pairs are characterized by larger mean infall velocities and velocity dispersions
with respect to their field-field counterparts. The exponential tails of their negatively skewed distribution are also much fatter.
In $\xi_{\rm s}$, halo-halo pairs produce the FoG enhancement at small $s_\perp$, while
field-field pairs completely dominate the signal at large $s$.

\begin{figure}
\centering
\includegraphics[scale=0.52]{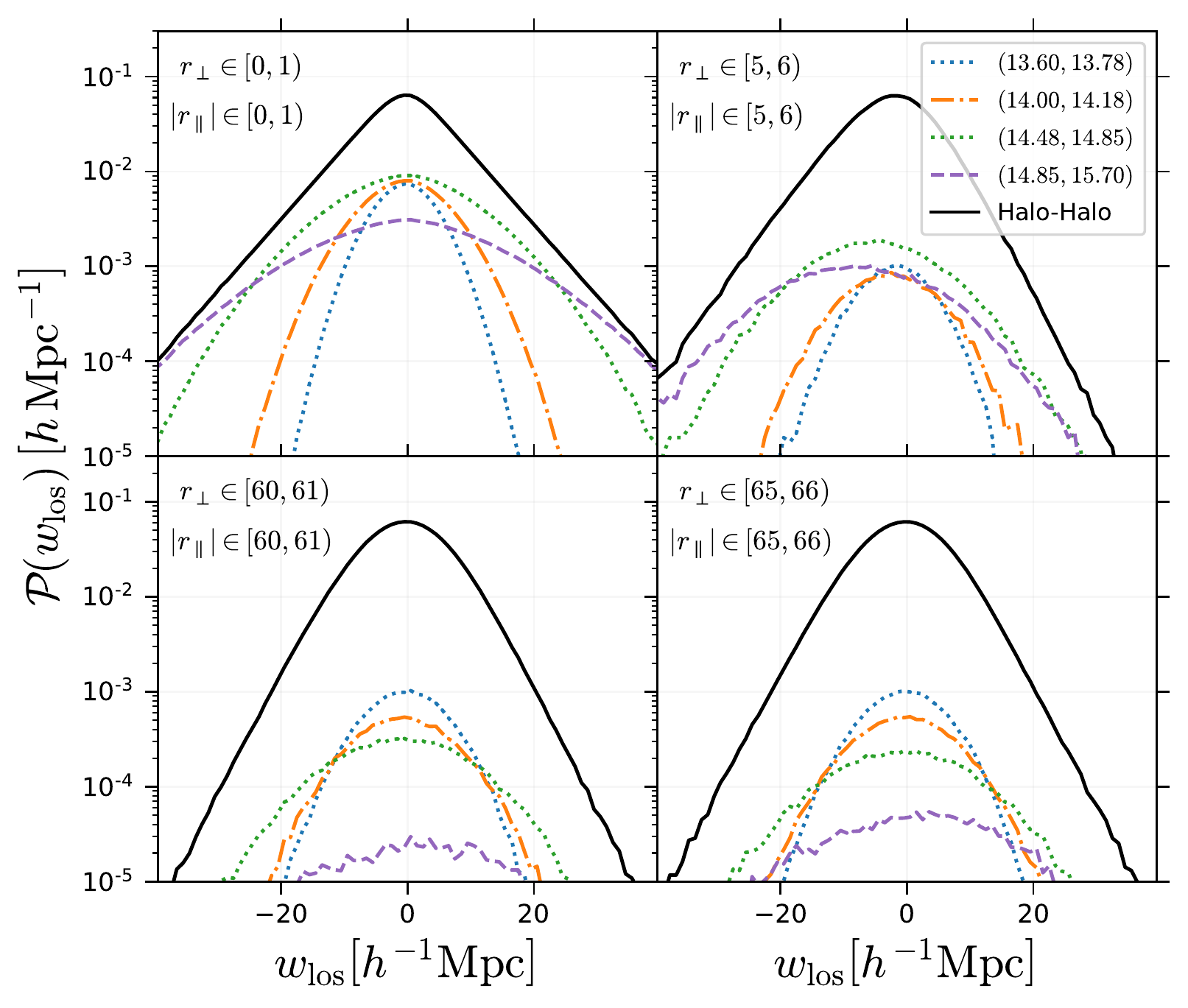}
\caption{The los pairwise-velocity distribution of the halo-halo term is further partitioned into the contributions of various halo log-mass bins (the bin boundaries in $\log_{10}[M_{\rm vir}/(h^{-1}\mathrm{M}_\odot)]$ are indicated in the labels) for
different real-space separation vectors (expressed in $h^{-1} \mathrm{Mpc}$). For simplicity, only `diagonal' terms in which both particles reside in haloes of the same bin are shown.} 
\label{fig:rel_los_comp_halos}
\end{figure}

In Fig.~\ref{fig:rel_los_comp_halos}, we further partition the halo-halo pairs into
subsamples based on the halo mass (for simplicity, cross pairs formed by particles in different halo-mass bins are not considered here). This procedure reveals that 
haloes of different masses\footnote{We also considered additional variables as the halo spin parameter
and triaxiality but their impact on $\mathcal{P}_{w_{\rm los}}$ was too small to be isolated.} are characterized by very different pairwise-velocity distributions. Not only the velocity dispersion increases with the halo mass but also
the mean infall velocity grows in magnitude. The skewness can even reverse sign for
cluster-sized haloes.
All this shows the complexity behind the overall distribution of $w_{\rm los}$ and clarifies why the PDF is so difficult to model
accurately.

\section{A new fitting function}
\label{sec:phe}
It has been proposed 
that $\mathcal{P}_{w_{\rm los}}(w_{\rm los}|{\mathbf{r}})$ 
can be more easily described as a superposition of simpler elementary functions.
The pairwise peculiar velocity dispersion depends on the (suitably defined) local density of the pairs
\citep{kepner+98} and its PDF can be modelled as a weighted sum of basic Gaussian terms evaluated at fixed halo masses and environment densities \citep{sheth96,sheth01,ti07_1,ti07_2}.
Integrating out the degrees of freedom due to the haloes, $\mathcal{P}_{w_{\rm los}}(w_{\rm los}|{\mathbf{r}})$
can be approximated as the
superposition of univariate Gaussian \citep{bi1} or quasi-Gaussian \citep{bi2} distributions whose cumulants are drawn from a multivariate Gaussian distribution.

Although relatively new in the field of cosmology, similar techniques have 
actually been in use for over a century in statistics, finance and the theory of turbulence. 
The basic idea is to suitably combine uncountably many elements of a parametric family of PDFs to model heavy tailed and skewed distributions.
In statistics,
an (uncountable) `mixture' or `compound' probability distribution is defined by the relation
\begin{equation}\label{eq:ansatz}
		\mathcal{P}_x(x|\tau) = \int \mathcal{P}_x(x|\mathbf{f})\, \mathcal{P}_{\mathbf{f}}(\mathbf{f}|\tau) \dif^n f = \mathcal{P}_x \circ \mathcal{P}_{\mathbf{f}}\;,
\end{equation}
where $x$ denotes the random variable of interest (subject to the condition $\tau$) and $\mathbf{f}$ is a $n$-dimensional
array containing the factors (or latent variables) that influence the distribution of $x$.
The function $\mathcal{P}_x(x|\mathbf{f})$ gives the conditional probability density
of $x$ in a subpopulation with given $\mathbf{f}$ while 
the `mixing distribution' $\mathcal{P}_{\mathbf{f}}$ is the joint probability density 
(or the statistical measure) of the factors.
If $\mathcal{P}_x(x|\mathbf{f})$ is a Gaussian distribution, $\mathcal{N}_x(x;\mu,\sigma^2)$,
then the variable $x$  
is called a mixture of Gaussians (or normals). 
Scale mixtures of normals (where mixing only involves $\sigma^2$) are widely used to model symmetric distributions with heavy tails.
Location mixtures of normals (where mixing only involves $\mu$) are commonly used to model
skewed\footnote{Note, however, that a location mixture with Gaussian mixing distribution yields another Gaussian.} distributions. 
Joint location and scale mixing gives rise to skewed distributions with heavy tails.

The above-mentioned models for $\mathcal{P}_{w_{\rm los}}(w_{\rm los}|{\mathbf{r}})$ can be phrased in this language. For instance, \citet{bi1} assume that $w_{\rm los}$ 
is a joint location and scale mixture of Gaussians with a bivariate Gaussian mixing distribution.
In this case, the scatter in $\mu$ and $\sigma$
is meant to represent physical variability due to some latent (but not-so-well-specified) environmental density. This method compresses the information contained in $\mathcal{P}_{w_{\rm los}}(w_{\rm los}|{\mathbf{r}})$ into five parameters (the mean values for $\mu$ and $\sigma$ and
the three elements of their covariance matrix) that change smoothly with $\mathbf{r}$.
Although the model matches well the outcome of numerical simulations on large scales, 
it has two drawbacks: 
i) by definition, $\sigma$ is non negative and thus cannot 
follow a Gaussian distribution\footnote{To address this issue, \citet{bi2} truncate the  $\sigma$ distribution at zero and renormalise it to account for the probability in the Gaussian tail with negative support.}; ii) the resulting $\mathcal{P}_{w_{\rm los}}(w_{\rm los}|{\mathbf{r}})$ cannot reproduce the large skewness measured at small separations ($r< 5-10\,h^{-1}$ Mpc depending on the redshift) for particles and haloes in $N$-body simulations \citep{bi2}. For this reason, \citet{bi2} replace the mixture of Gaussians
with a mixture of skewed quasi-Gaussians obtained by applying the Edgeworth expansion
to first order. However, in order to limit the number of degrees of freedom of the model, the skewness of $\mathcal{P}_x(x|\mathbf{f})$ is not used as a third factor over which
the mixing is performed but is deterministically linked to the variances of $\sigma$ and
$\mu$ by means of an ansatz that introduces an additional free parameter. 

While these recent efforts have led to the development of tools that 
can closely approximate $\mathcal{P}_{w_{\rm los}}(w_{\rm los}|{\mathbf{r}})$ on a wide
range of scales, they are based on to a phenomenological description
that provides little insight into the underlying physics.
At the current stage of development, the models do not have predictive power.
They essentially offer a language and a convenient class of fitting functions
that can be used to describe the output of simulations with different gravity models and retrieve information on the cosmological parameters
and the law of gravity from the
observed $\xi_{\rm s}$ (assuming a functional form for the scale dependence of the model parameters).
Their complexity, however, is growing rapidly and, as we discussed above, ad hoc assumptions are required to limit their degrees of freedom while
extending their range of validity. 
Given this premise, we follow here a pragmatic and complementary approach by proposing
a fitting function for $\mathcal{P}_{w_{\rm los}}(w_{\rm los}|{\mathbf{r}})$ 
that closely reproduces the features seen in $N$-body simulations and provides an excellent fit to them at all scales. 
For this purpose, we search the statistics literature for a family of analytic PDFs with the following characteristics:
i) unimodality;
ii) presence of quasi-exponential tails; iii) highly tunable low-order cumulants (in particular skewness); iv) possibility of reducing to the Gaussian distribution in some limit.
We end up selecting the generalised hyperbolic distribution (GHD) which will be precisely defined in the next section. 
As in \citet{Peebles76} and
\citet{reid11},
we do not give any physical motivation in support of our choice but we note that,
interestingly enough, the GHD describes a particular mixture of normals. 

\subsection{The generalised hyperbolic distribution}
\label{sec:GHD}
\subsubsection{The inverse Gaussian distribution}
\label{sec:IG}
Let us consider a one-dimensional standard Wiener process with drift $\nu \in \mathbb{R}_{>0}$ and diffusion coefficient $\sigma \in \mathbb{R}_{>0}$. At time $t$, the position $x$ of a
Brownian particle follows the distribution $\mathcal{N}_x(x;\nu t, \sigma^2 t)$.
The first-passage time of the level $\ell>0$ by a Brownian walker is distributed as\footnote{In cosmology, this PDF has been used by \citet{Bond91} to solve the cloud-in-cloud problem in the excursion-set model for the halo mass function.} \citep{Schroedinger1915}
\begin{equation}
f_t(t;\ell,\nu,\sigma)=\frac{\ell}{\sqrt{2 \pi} \sigma}\, t^{-3/2}\,\exp{\left[-\frac{(\ell-\nu t)^2}{2\sigma^2 t}\right]}\;.
\end{equation}
When expressed in terms of the parameters $\mu=\ell/\nu$ and $\gamma=\ell^2/\sigma^2$,
this equation defines the `inverse Gaussian distribution' \citep[see e.g.][for a comprehensive review of its properties]{chhikara-folks-book-1989, seshadri-book},
\begin{equation}
\mathcal{I}_t(t;\mu,\gamma)=\left( \frac{\gamma}{2 \pi } \right)^{1/2}\, t^{-3/2}\,\exp{\left[-\frac{\gamma\,(t-\mu)^2}{2\mu^2 t}\right]}\;.
\label{eq:ig}
\end{equation}
This PDF provides a classic model for non-negative, unimodal and positively skewed data and is widely employed to perform lifetime and survival studies in ecology, engineering, finance, law and medicine.
The mean of the distribution coincides with the location parameter $\mu$,
while the variance ($\mu^3/\gamma$), skewness ($3\sqrt{\mu/\gamma}$) and kurtosis
($15\mu/\gamma$) also depend on the scale parameter $\gamma$.
The name inverse Gaussian was coined by \citet{Tweedie56} and only
refers to the fact that, while the Gaussian distribution describes the distribution of $x$ at fixed $t$, $\mathcal{I}_t$ gives the PDF of the time at which the particles first cross a fixed position.

\subsubsection{Generalised inverse Gaussian distribution}
In the 1940s and 1950s, a larger family of unimodal PDFs with positive support
was introduced. Since this class includes the inverse Gaussian distribution as a special case, it now goes under the name of the `generalised inverse Gaussian distribution' \citep{jorgensen82}. 
The corresponding PDF for the random variable $t>0$ is
\begin{equation}
	\mathcal{G}_t(t; \lambda, \chi, \psi) = \frac{(\psi/\chi)^{\lambda/2}}{2\,K_{\lambda}(\sqrt{\psi\chi})}\, t^{\lambda-1} \,\exp\left[-\frac{1}{2}\,\left(\psi t+\frac{\chi} {t}\right)\right] \; ,
    \label{eq:gig}
\end{equation}
where $\lambda \in \mathbb{R}$, $\chi \in \mathbb{R}_{>0}$, $\psi \in \mathbb{R}_{>0}$ and $K_{\lambda}$ is the modified Bessel function of the second kind with order $\lambda$ \citep{as72}.
For $\lambda=-1/2$, $\psi=\gamma/\mu^2$ and $\chi=\gamma$,
equation (\ref{eq:gig}) reduces to the inverse Gaussian distribution given in equation
(\ref{eq:ig}).
The generalised inverse Gaussian distribution can be interpreted as the distribution of the first passage time or, depending on the sign of $\lambda$, the last exit time of more complicated diffusion processes than Brownian motion \citep{Vallois91}.
Note that, for large values of $t$, $\mathcal{G}_t$ has a thin tail $\propto t^{\lambda-1} \,\exp(-\psi t/2)$.
Another interesting property is that, if $t$ follows $\mathcal{G}_t(t;\lambda, \chi, \psi)$,
then $t^{-1}$ follows $\mathcal{G}_{t^{-1}}(t^{-1};-\lambda, \psi, \chi)$.
In analogy with $\mathcal{I}_t$, the generalised inverse Gaussian distribution finds many direct applications in risk assessment and queue modelling. 
Moreover, it is commonly used as a mixing distribution whenever there is the need for
skewed weighting.
This practice was initiated by \citet{Sichel74} who used mixtures of Poisson distributions 
to model the distribution of sentence lengths and word frequencies.
   	\begin{figure*}
	\centering
	\includegraphics[scale=0.7]{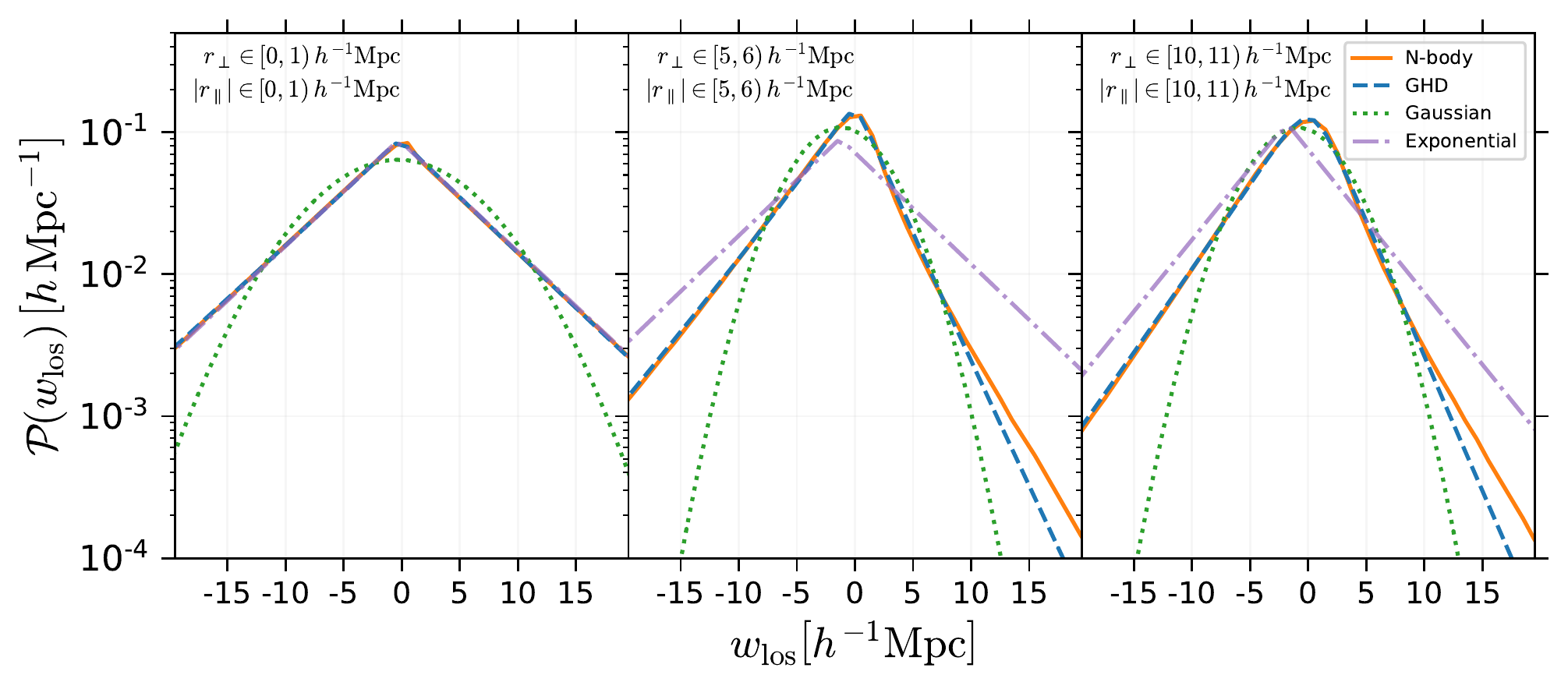}
	\caption{The los pairwise velocity distribution for the DM particles in the \textsc{W0} simulation at $z=0$ (solid) is compared with the best-fitting GH (dashed), exponential (dot-dashed) and Gaussian (dotted) approximations for three different pair-separation distances in real space.}
	\label{fig:gh}
	\end{figure*}		

\subsubsection{Normal variance-mean mixtures}
Let us return to the Wiener process with drift we introduced in Section \ref{sec:IG}. 
This time, however, we assume that the Brownian particles start from $x=\alpha\in \mathbb{R}$
at $t=0$. At a given time $t>0$, the position of a random Brownian walker is then $x=\alpha+\nu t+\sigma \sqrt{t}\,g$ where $g$ is a Gaussian random variable with zero mean and unit variance.
Let us now introduce a second random variable $p>0$ which is independent of $g$ and follows a generic distribution $\mathcal{P}_p$. We use $p$
to pick random times at which we sample the positions of the Brownian particles. This
leads us to consider the random variable
$x=\alpha+\beta p+\sigma \sqrt{p}\,g$ which is a non-linear combination of $g$ and $p$. 
Its PDF, 
\begin{equation}
\mathcal{P}_x(x)=\mathcal{N}_x(x;\alpha+\beta\,p,\sigma^2\,p)\circ \mathcal{P}_p(p)\;,
\end{equation}
is called a normal variance-mean mixture \citep{Barndorff-Nielsen-Kent-Sorensen-82}.
A theorem shows that if $p$ is unimodal then so is $x$ \citep{Yu11}.

		\begin{figure*}
		\centering
        {\includegraphics[width=0.33\textwidth]{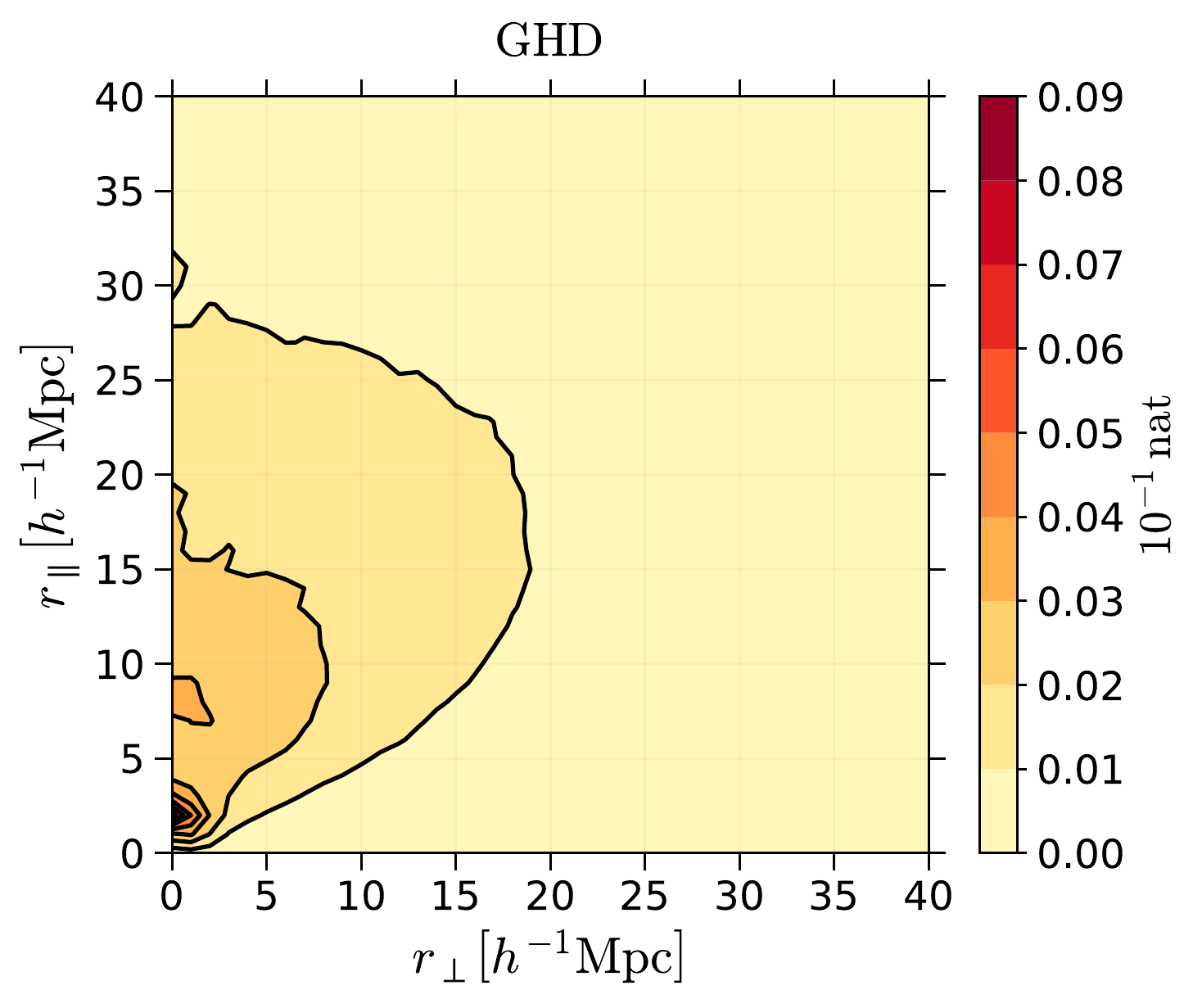}}%
	{\includegraphics[width=0.33\textwidth]{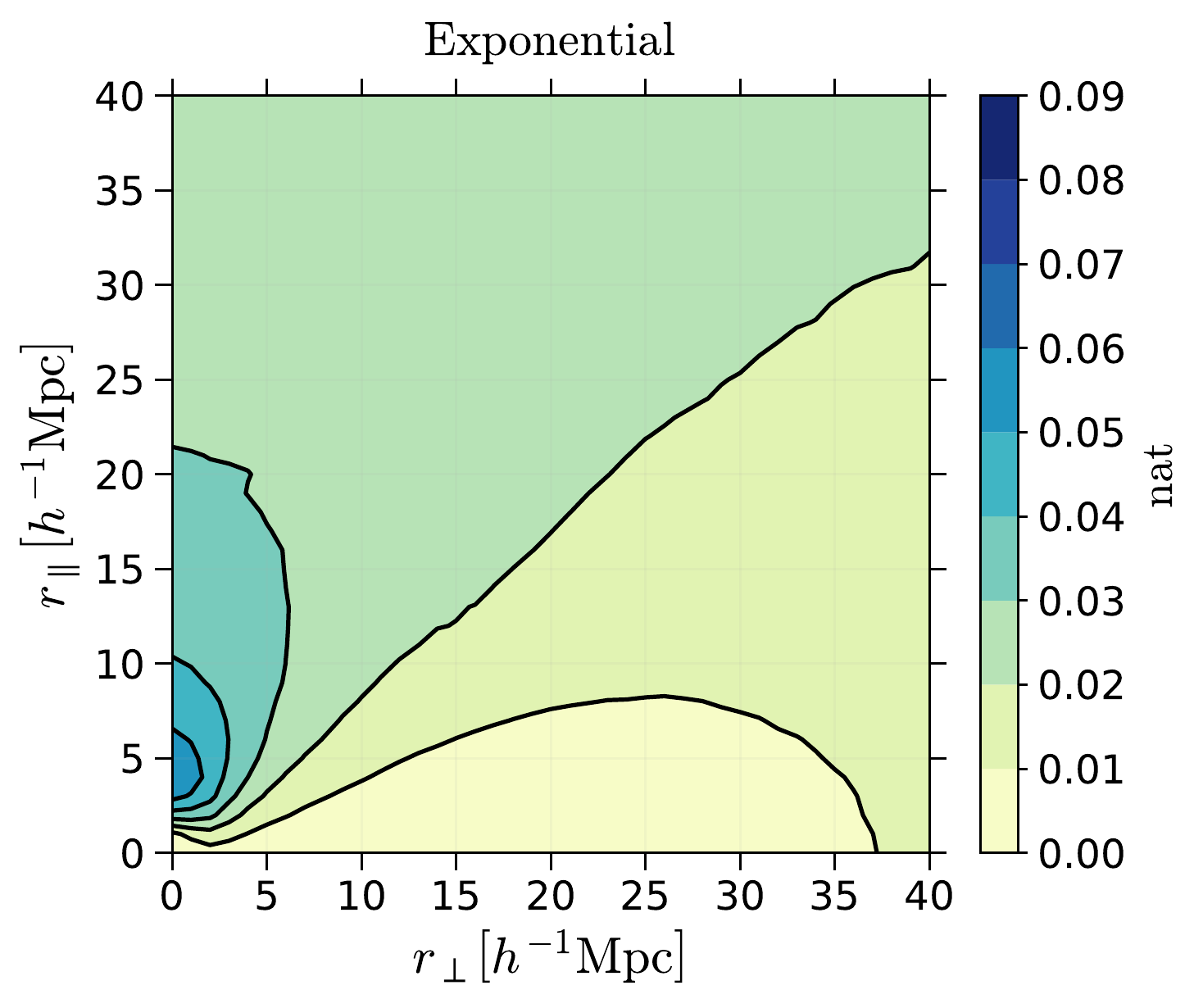}}%
	{\includegraphics[width=0.33\textwidth]{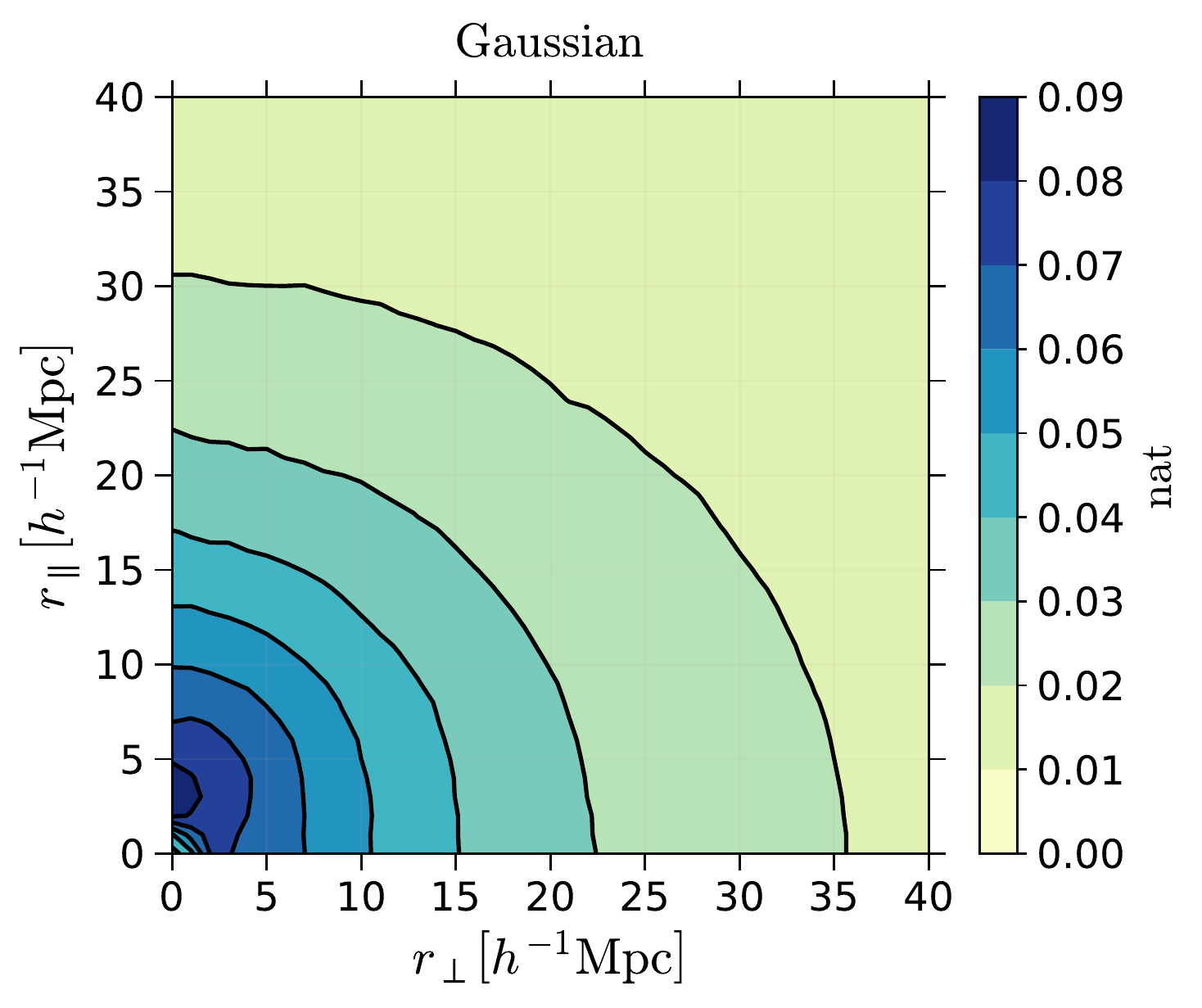}}
		\caption{KL divergence for the GHD (left), exponential (middle) and Gaussian (right) best-fitting functions
        with respect to the los pairwise velocity distribution measured in the simulation. The reported values are expressed in natural units of information. Note that the color bar in
        the left panel is compressed by a factor of 10 compared with the middle and right panel.}
		\label{fig:KL}
	\end{figure*}

\subsubsection{The generalised hyperbolic distribution}
A Gaussian PDF plotted on a semi-logarithmic graph describes a parabola. 
Although the frequency distribution of many empirical phenomena shows this property, 
there exist cases in which an hyperbola provides a much better description than a parabola
due to the presence of exponential tails.
A classic example is the log-size distribution of sand grains in natural aeolian deposits \citep{Bagnold41} and many others arise particularly in finance.
The name `hyperbolic distributions' has been coined to designate this class of probability
densities. The very first analytical example of such a PDF was derived in physics
by calculating the distribution of particle velocities in an ideal relativistic gas
\citep{Juettner-1911}.

The GHD is a larger family of PDFs that includes the hyperbolic distributions
as a particular case.
It was introduced by \citet{BN77} in order to model the log-size distribution of sand grains
and is defined as 
a normal variance-mean mixture in which
$\sigma=1$ and $\mathcal{P}_p(p) = \mathcal{G}_p(p;\lambda, \delta^2, \alpha^2-\beta^2)$.
Its PDF 
takes the form \citep{prause99}
\begin{align}
&\mathcal{H}_x(x; \alpha, \beta, \delta, \lambda, \mu)  = \mathcal{N}_x(x; \mu + \beta\,p, p)\circ\mathcal{G}_p(p; \lambda, \delta^2, \sqrt{\alpha^2-\beta^2})= \nonumber \\
&\,C\,
\left[\delta^2+\left(x-\mu\right)^2\right]^{\frac{\lambda-1/2}{2}}
e^{\beta(x-\mu)}\,
K_{\lambda-\frac{1}{2}}\left(\alpha\,\sqrt{\delta^2+(x-\mu)^2}\right) \; ,
\end{align}
\noindent with normalisation constant
	\begin{equation}
		C
        = \frac{\left(\alpha^2-\beta^2\right)^{\frac{\lambda}{2}}}{\sqrt{2\pi}\,\alpha^{\lambda - 1/2}\,\delta^\lambda\,K_{\lambda}\left[\delta\,\sqrt{\alpha^2-\beta^2}\right]}  \; .
	\end{equation}
The domain of variation of the parameters is 
	\begin{align}
	\delta \geq 0, \ \ \abs{\beta} < \alpha, \;\;\;\; \mathrm{if}\,\, \lambda > 0 \; , \nonumber \\ \label{eq:domain}
	\delta > 0, \ \ \abs{\beta} < \alpha, \;\;\;\; \mathrm{if}\,\, \lambda = 0 \; , \\
	\delta > 0, \ \ \abs{\beta} \leq \alpha, \;\;\;\; \mathrm{if}\,\, \lambda < 0 \; . \nonumber    
	\end{align}
It is not easy to isolate the impact of each of them and several
alternative parameterizations of the GHD have been introduced to alleviate this problem.
Broadly speaking, $\lambda$ defines various subclasses and influences the tails, $\alpha$ modifies the shape (i.e. variance and kurtosis), $\beta$ the skewness, $\delta$ the scale and $\mu$ shifts the mean value.
A convenient property of the GHD is that it reduces to several named distributions in the appropriate limit. For example, it gives the hyperbolic distribution for $\lambda=1$ and 
becomes a Gaussian distribution with variance $\delta/\alpha$ when both $\alpha$ and $\delta$ tend to infinity \citep{titel10}.

The GHD shows semi-heavy tails \citep{BN+Blaesild-81},
\begin{equation}
	\mathcal{H}_x \sim \abs{x}^{\lambda-1}\,\exp(-\alpha \abs{x}+\beta x)\,\,\,\,\,\,\,\mathrm{as}\,\,\,\,\,\,\,x\rightarrow \pm\infty\;,
\end{equation}
and all its moments exist.
The moment generating function is
	\begin{equation}
	M(x) = e^{\mu x}\,\left[\frac{\alpha^2-\beta^2}{\alpha^2-(\beta+x)^2}\right]^{\lambda/2} \frac{K_{\lambda}\left(\delta\sqrt{\alpha^2-(\beta+x)^2}\right)}{K_{\lambda}\left(\delta\sqrt{\alpha^2-\beta^2}\right)} \, ,
	\end{equation} 
with $\abs{\beta+x} < \alpha$ which follows from equation (\ref{eq:domain}). 
Explicit expressions for the first four moments and cumulants are given in
\citet{BN+Blaesild-81}.

	\begin{figure}
	\centering
	\includegraphics[scale=0.65]{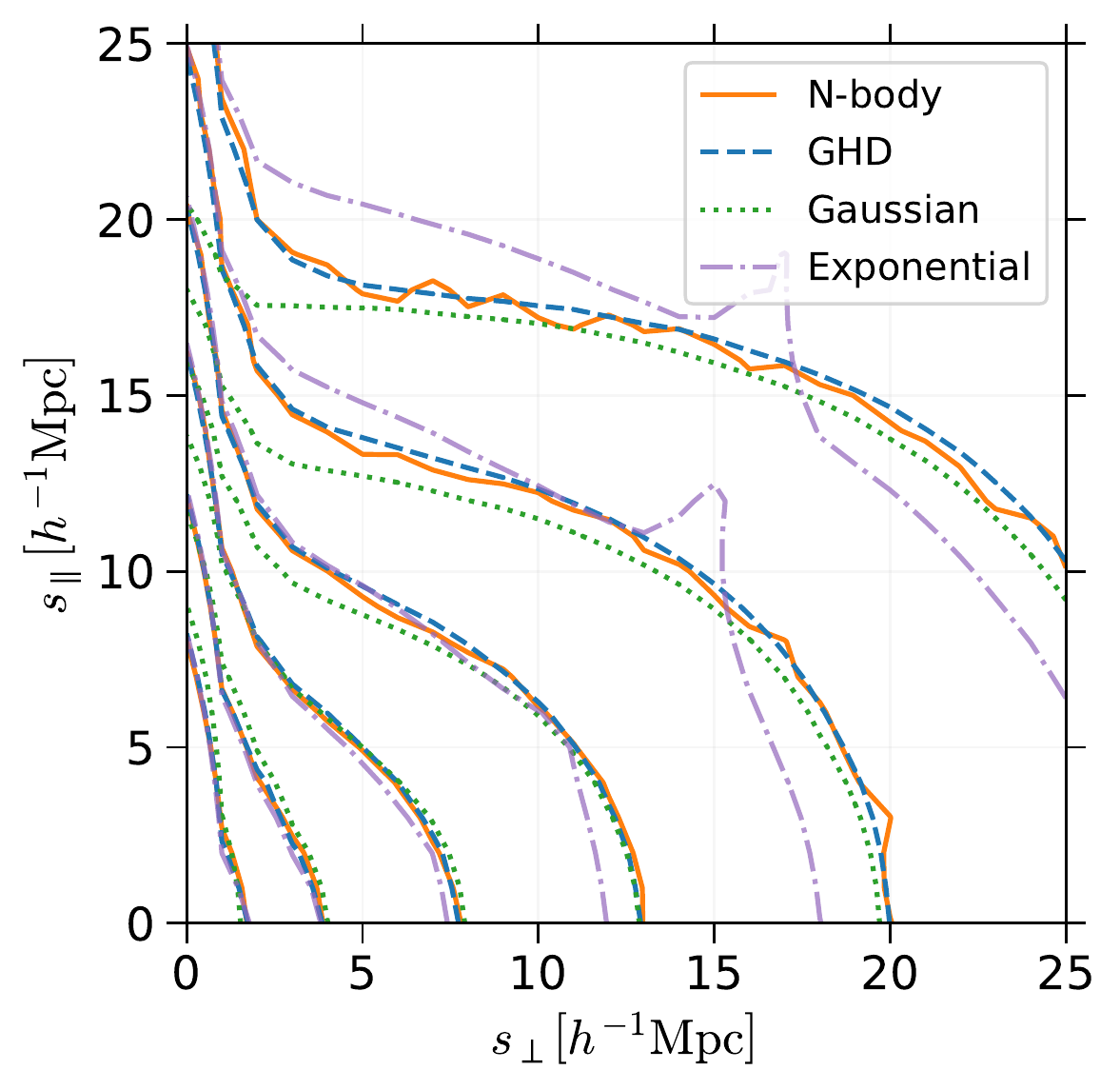}
	\caption{The redshift-space correlation function $\xi_{\rm s}$ for the particles in our $N$-body simulations (solid) is compared with the outcome of the streaming model obtained
    by fitting either a GHD (dashed), a Gaussian distribution (dotted) or an exponential (dash-dotted) to $\mathcal{P}_{w_{\rm los}}(w_{\rm los}|\mathbf{r})$. }
	\label{fig:corr_gh}
	\end{figure}

		\begin{figure*}
		\centering
        {\includegraphics[width=0.77\textwidth]{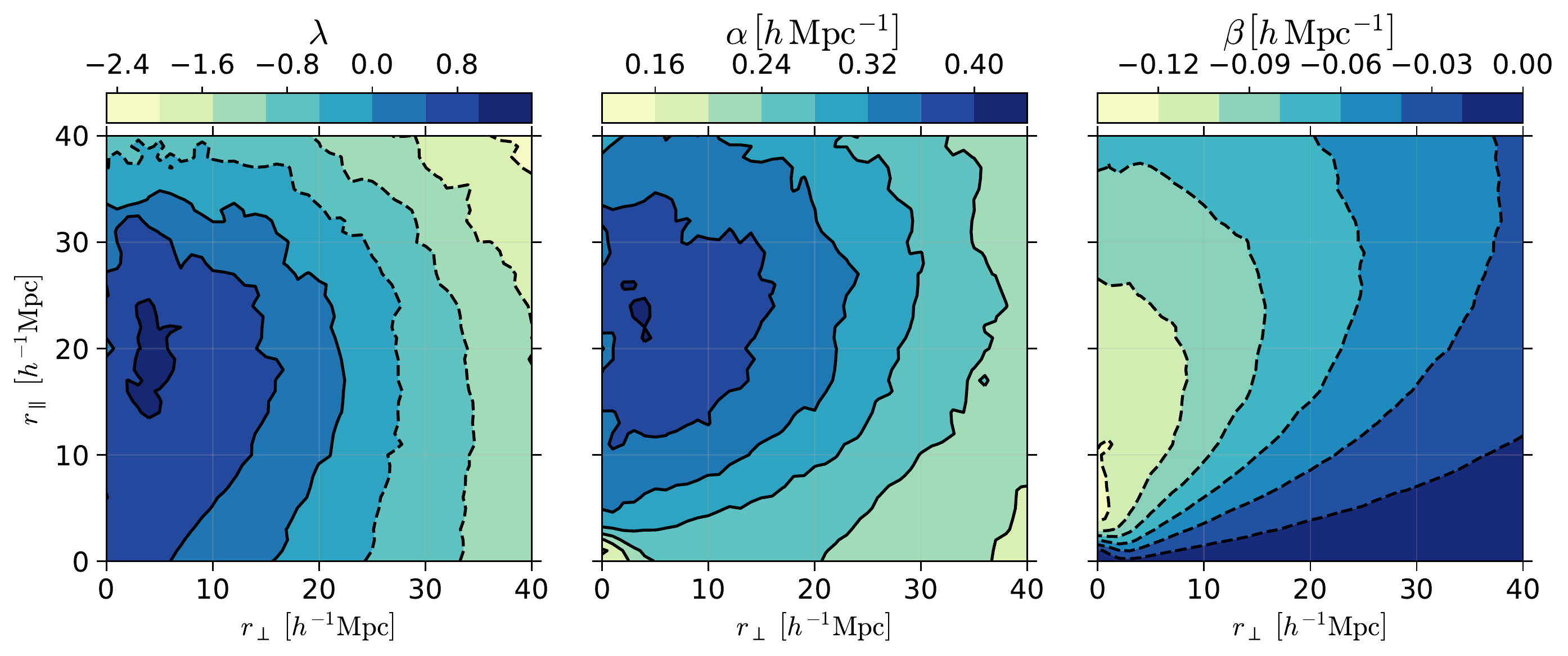}}%
		\hfill
	{\includegraphics[width=0.52\textwidth]{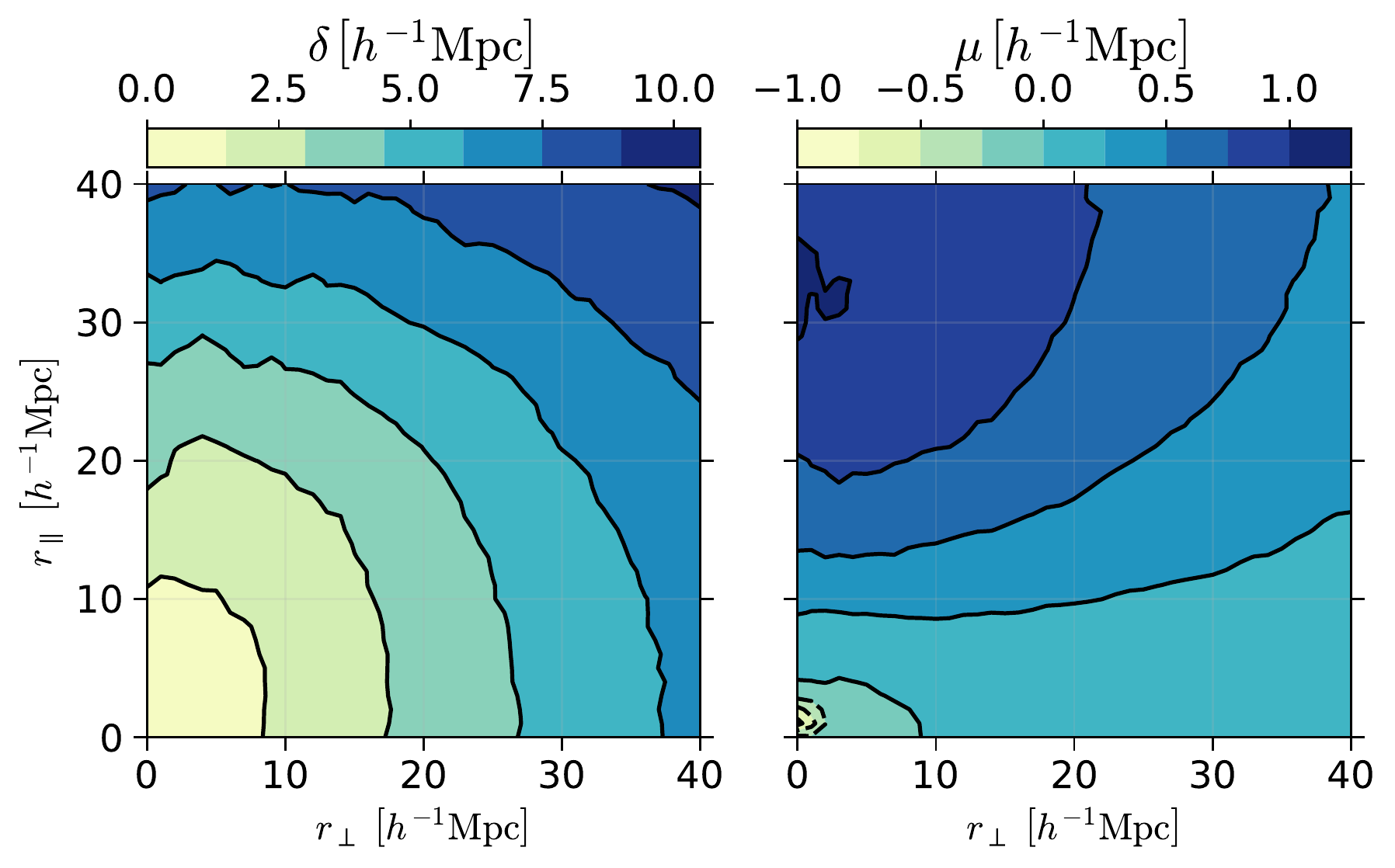}}
		\caption{Best-fitting parameters of the GHD as a function of $r_\parallel$ and
        $r_\perp$.}
		\label{fig:GHDparams}
	\end{figure*}
 
\subsection{Application to pairwise velocities}
\subsubsection{DM particles}
In Fig.~\ref{fig:gh}, we show that the GHD (dashed line) provides a very good fit to the histogram of $w_{\rm los}$ extracted from the W0 simulation (solid line). 
The optimal values for the parameters
have been determined assuming (symmetrised) Poisson errors \citep{Gehrels86} and using least-squares fitting with a Markov Chain Monte Carlo sampler (we have checked for a few separations that this method gives consistent results with a maximum-likelihood analysis which is time consuming given the huge number of particle pairs).
Our results show that
the GHD accurately describes $\mathcal{P}_{w_{\rm los}}$ around the mode and in the negative
tail while it slightly underestimates the PDF in the positive tail. The improvement with
respect to Gaussian fits (dotted lines) is dramatic as the normal distribution cannot
match the exponential tails seen in the simulation.
For very small spatial separations, the exponential distribution given in equation (\ref{eq:expfull})
also provides a very good fit (dot-dashed lines). However, the agreement with the simulation
data rapidly decreases with increasing $r$ as the model has the wrong shape
around the mode of the distribution.
A more quantitative analysis is performed in Fig.~\ref{fig:KL} where we compare
the information loss associated with the GHD, exponential and Gaussian approximations.
Shown is the Kullback-Leibler (KL) divergence between the actual PDF measured in the
simulations and the three approximations as a function of the real-space separation of the pairs. This quantity provides a
measure of goodness of fit. 
For the range of separation vectors shown in Fig.~\ref{fig:KL},
the information loss associated with the Gaussian approximation is always at least one
order of magnitude larger than for the GHD. This property persists also on larger scales.
Similar conclusions can be drawn comparing
the exponential and the GHD approximations, although, in this case, the fits are of similar quality at very small separations ($r_\parallel<1\,h^{-1}$ Mpc and $r_\perp<15\,h^{-1}$ Mpc).
Note that the GHD compression is nearly lossless at all scales.

Finally, in Fig.~\ref{fig:corr_gh} we show the redshift-space correlation function
obtained with the streaming model by inserting the best-fitting GHD in equation
(\ref{eq:streaming}) together with the real-space correlation function extracted
from the simulation. Our results (dashed lines) provide an excellent description
of the redshift-space correlation measured in the simulation (solid lines).
For $s>5\,h^{-1}$ Mpc, deviations are comparable with the Poisson error 
for $\xi_{\rm s}$ which is always between one and a few per cent.
On smaller scales, where the Poisson error becomes substantially sub per cent, one starts noticing statistically significant deviations at a few per cent level
(not visible in the plot).
For comparison, we also show the results obtained using the Gaussian and exponential fits for $\mathcal{P}_{w_{\rm los}}$. The Gaussian model (dotted lines) shows large systematic deviations on small scales and matches the simulations (at the level of the Poisson errors) only for $s_\parallel> 20\,h^{-1}$ Mpc and $s_\perp> 5\,h^{-1}$ Mpc. 
The exponential model\footnote{The features that are noticeable in the contour map of $\xi_{\rm s}$ at $s_\parallel\simeq s_\perp$ are caused by discontinuities in $\bar{w}_{\parallel}$ and $\sigma_{12}$ as a function of $r_\parallel$ and $r_\perp$. In fact, the posterior probability density of these parameters is bimodal for a range of real-space separations and we selected the peak with the highest integrated probability to draw the dot-dashed lines in Fig.~\ref{fig:corr_gh}.} (dot-dashed lines), on the other hand, is accurate only at small $s_\perp$. 

\begin{figure*}
		\centering
        {\includegraphics[width=0.6\textwidth]{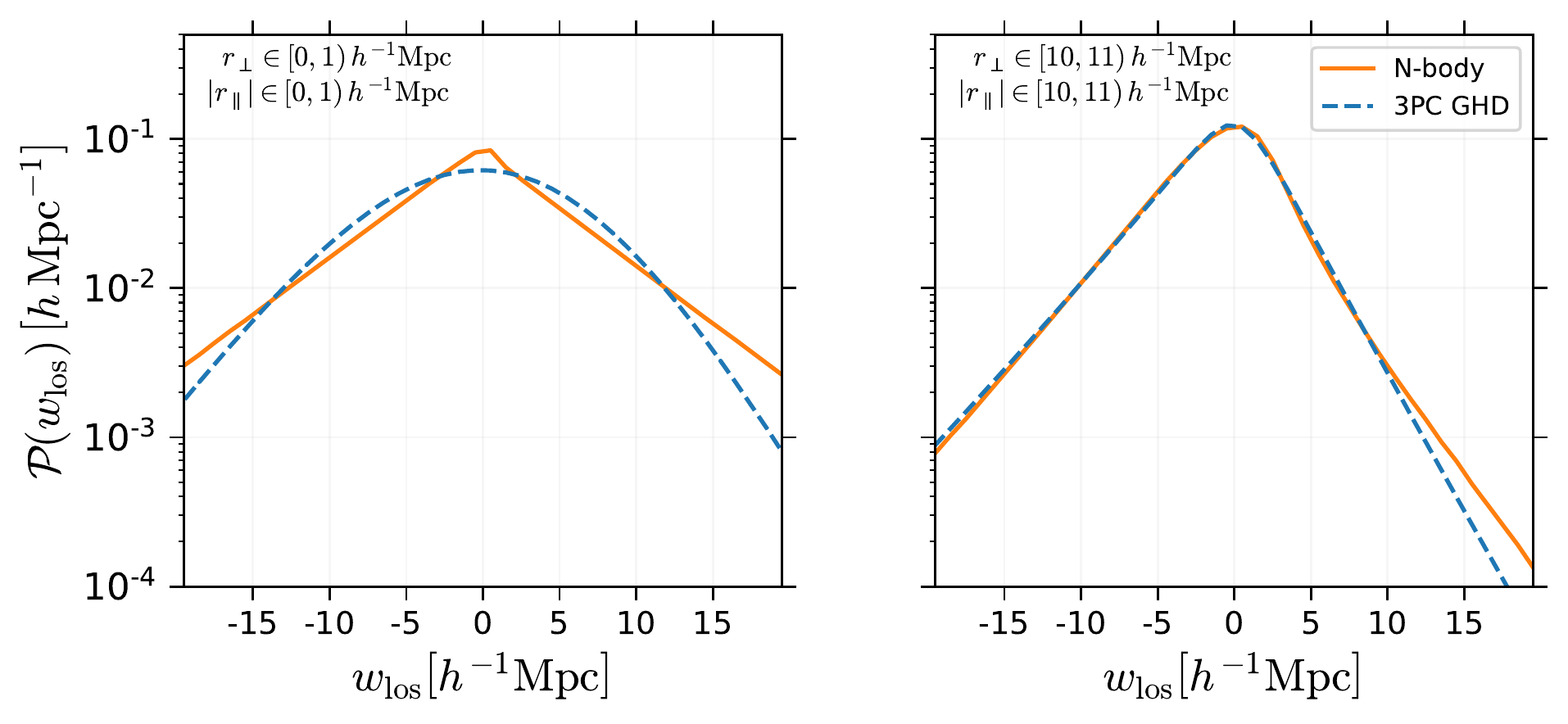}}%
		\hfill
	{\includegraphics[width=0.34\textwidth]{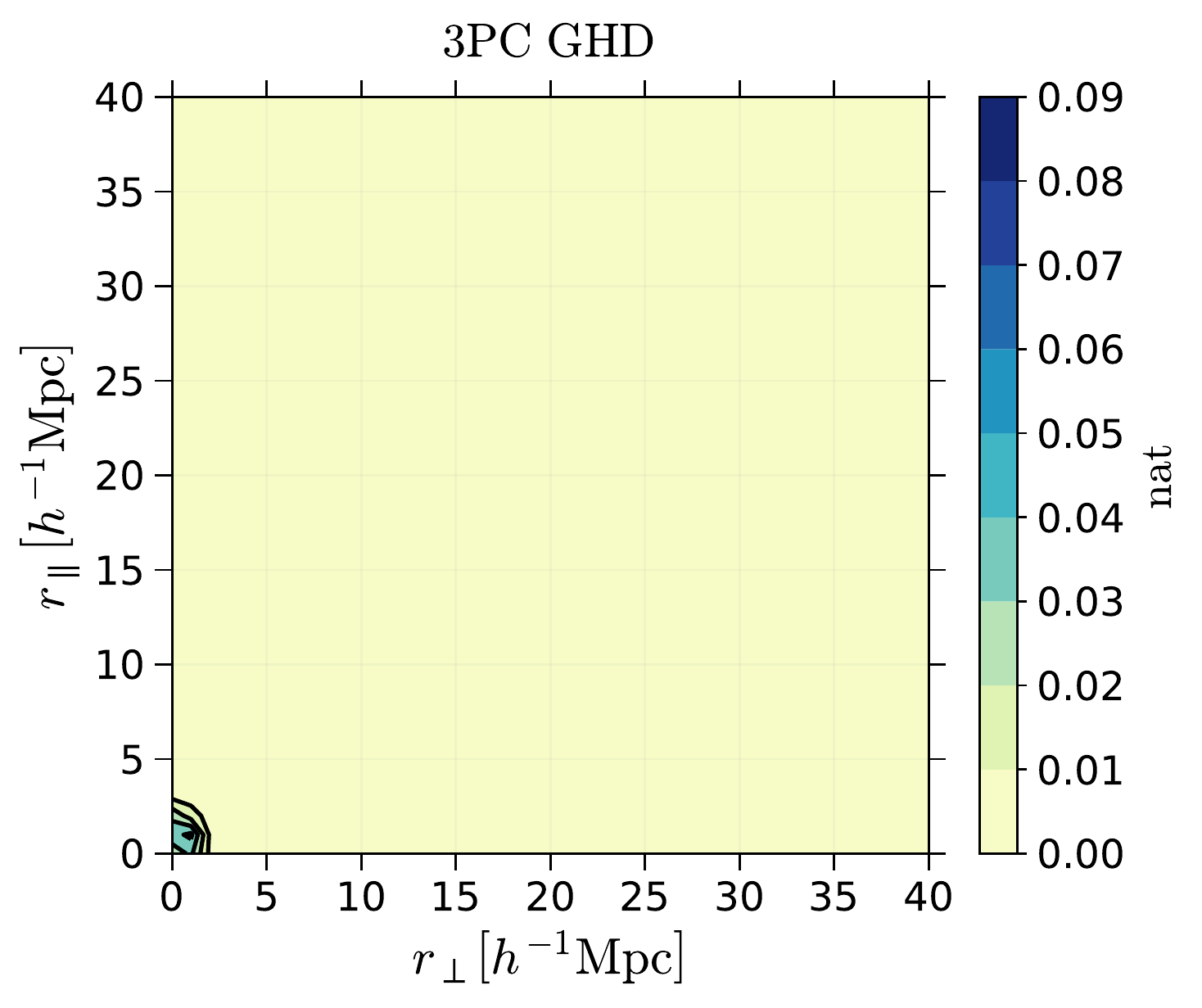}}%
	{\includegraphics[width=0.29\textwidth]{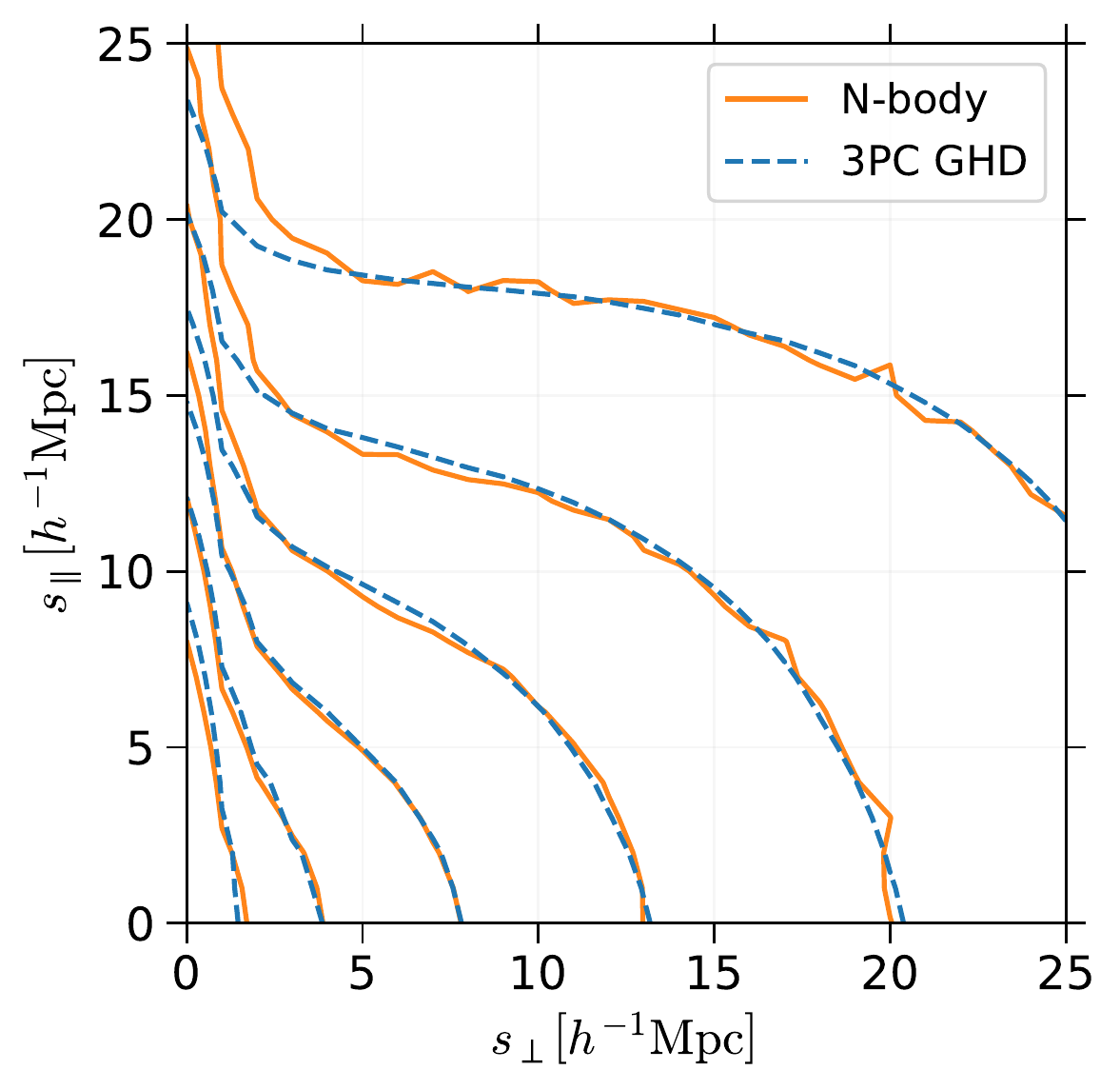}}
		\caption{As in Figs. \ref{fig:gh}, \ref{fig:KL} and \ref{fig:corr_gh} but for a GHD
with only 3 degrees of freedom that identify the position of the model parameters  
within the volume spanned by the first 3 principal components. }
		\label{fig:pca}
	\end{figure*}

The high fidelity of the GHD fit comes at the price of using five free parameters. 
Their marked scale dependence (see Fig.~\ref{fig:GHDparams}) represents a severe limitation for future practical applications
that aim to interpret observational data.
We note, however, that the best-fitting parameters at different $\mathbf{r}$ tightly cluster along a flattened sequence in five-dimensional space. By applying a principal component analysis to the standardised variables, we find that the first three components account for 99.8 per cent of the variance. We thus fit again the distribution of $w_{\rm los}$ in the simulation using only 
three free parameters that denote the position of $\alpha, \beta, \delta, \lambda$ and $\mu$ within the space spanned
by the first three principal components. In Fig.~\ref{fig:pca}, we show the quality of the best-fitting functions as well as the corresponding $\xi_{\rm s}$ obtained by inserting them into equation~(\ref{eq:streaming}).
The three-parameter GHD still outperforms the Gaussian approximation at all scales. 
 	\begin{figure}
		\centering
        { \includegraphics[scale=0.52]{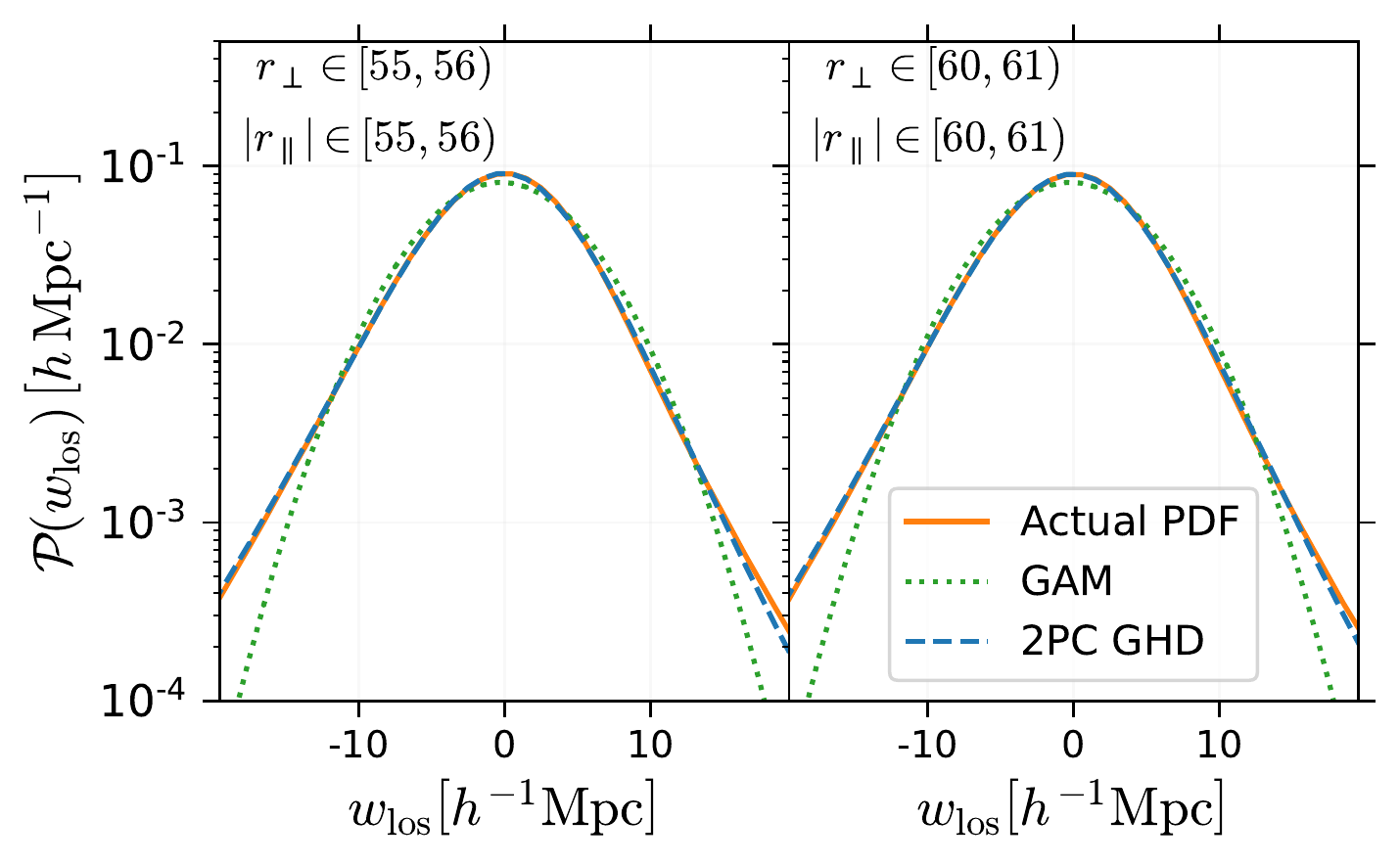}       }%
		\vfill
	{\includegraphics[width=0.42\textwidth]{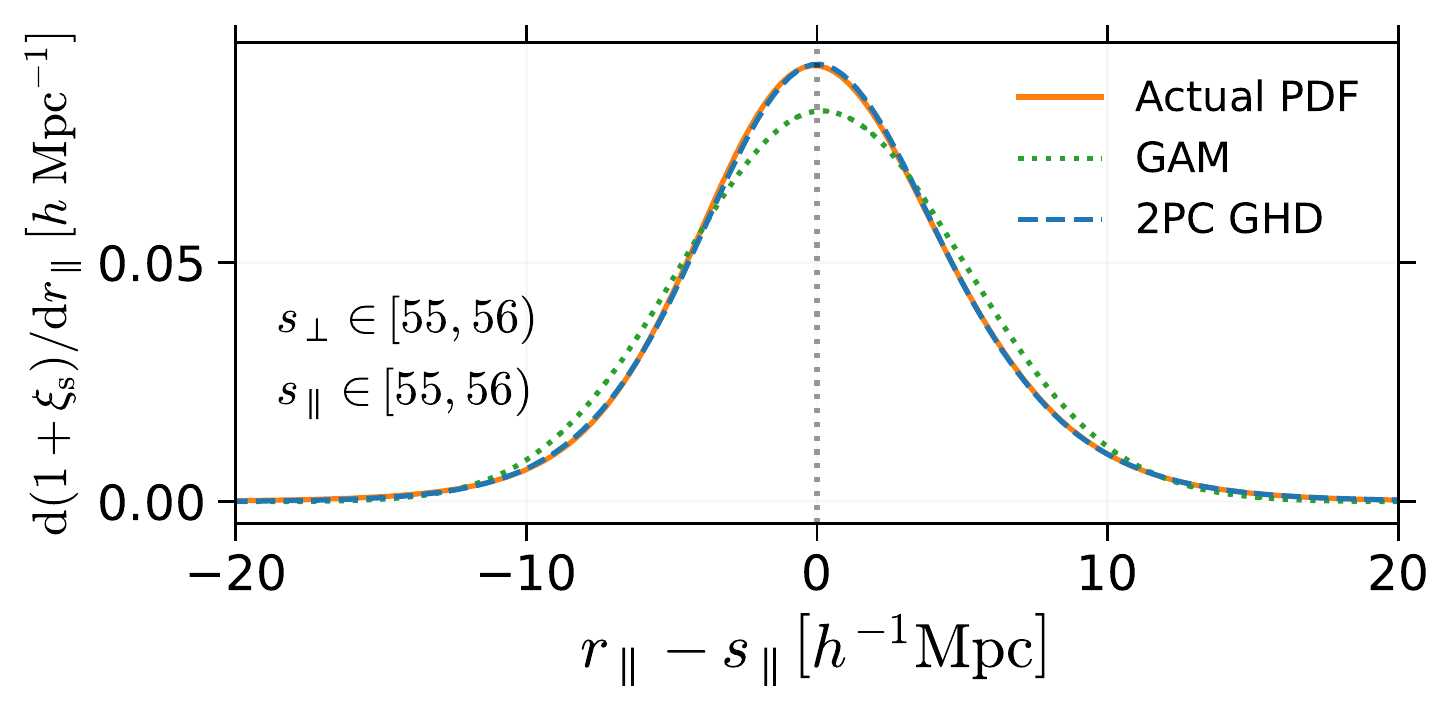}}%
		
		\caption{A simplified GHD model for $\mathcal{P}_{w_{\rm los}}$
in which the number of free parameters is reduced to 2 through PCA (2PC GHD, dashed) provides an excellent description of the $N$-body data (solid) on large scales and vastly outperforms the Gaussian approximation (dotted).
In the top panel, we directly  
compare data and best-fit models for the pairwise-velocity PDF (as in Fig.~\ref{fig:gh} but for larger pair separations given in units of $h^{-1}$ Mpc). 
In the bottom panel, we show the corresponding integrand of the streaming equation, as in Fig.~\ref{fig:clpt_xis}.} 
		\label{fig:reducedGH}
	\end{figure}		   
This requirement can be relaxed at larger scales where $\mathcal{P}_{w_{\rm los}}$ 
assumes a simpler shape. Fig.~\ref{fig:reducedGH} shows that a two-parameter GHD
obtained through PCA for larger scales
provides an excellent fit that better describes the tails of the distribution with
respect to the Gaussian approximation at all spatial separations. 
Note that even the integrand
of the streaming equation is impeccably
reproduced in this case.

 \begin{figure*}
	\centering
	\includegraphics[scale=0.75]{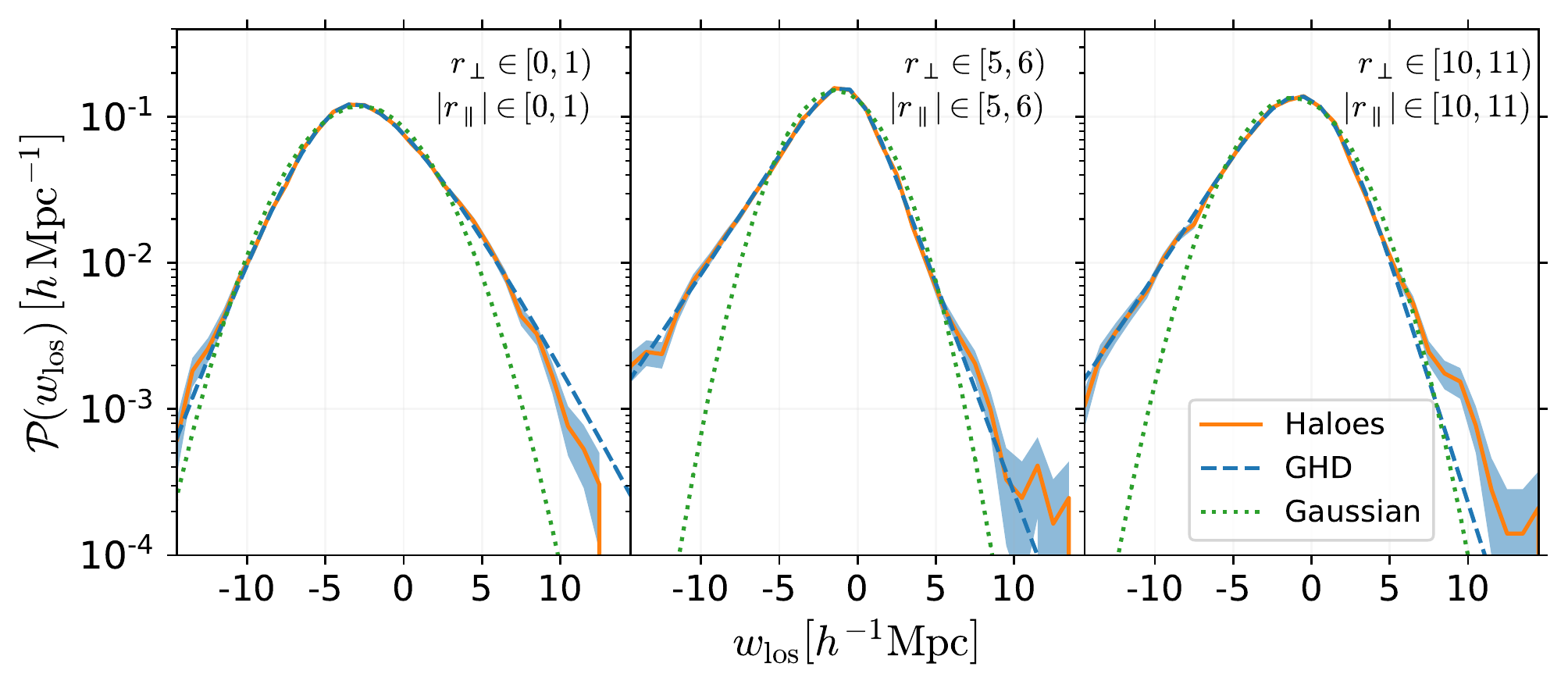}
	\caption{The los pairwise velocity distribution for the DM haloes identified in the W0 simulation at $z=0$ (solid line with shaded error) is compared with the best-fitting GH (dashed) and Gaussian (dotted) approximations for diffferent
 real-space pair separations expressed in 
$h^{-1} \mathrm{Mpc}$.}
	\label{fig:GHD_halo}
\end{figure*}

\subsubsection{DM haloes}
\label{app:1}

DM haloes are biased tracers of the matter distributions and, as discussed in Section \ref{halobias}, are also characterized by a different pairwise-velocity PDF which presents less
prominent tails than for the matter. In fact, haloes are not subject to the Finger-of-God effect and have a substantially smaller pairwise velocity dispersion,
especially at small separations.
In Fig.~\ref{fig:GHD_halo}, we 
show that the GHD provides an excellent fit to
the halo pairwise velocity distribution and, also in
this case, outperforms the Gaussian approximation.

\subsection{Discussion}
\subsubsection{Comparison with previous work}
 The GHD is a mixture of Gaussians analogous to that introduced in \citet{bi1}, although with a very different mixing
 distribution.
 It is thus interesting to highlight similarities and differences between the two approaches.
In both cases, skewness is generated by the correlation between the mean and the variance of the constituent Gaussian distributions. However, the two models achieve this differently.
In \citet{bi1}, $\mu$ and $\sigma$ are Gaussian random variables and their correlation is a free parameter.
The GHD instead originates from the deterministic relation  $\mu=\alpha+\beta\sigma^2$.
Additional skewness appears in the GHD because the mixing distribution $\mathcal{G}$ is itself skewed. This illustrates why the GHD can accommodate the large skewness measured
in $N$-body simulations at small $r$ while the model by \citet{bi1} cannot \citep[see][]{bi2}.

Another difference between the two approaches lies in the support of the mixing distribution. 
Although the rms value $\sigma$ is by definition non-negative,
\citet{bi1} use a mixing function with support on $\mathbb{R}^2$ for $\mu$ and $\sigma$.
The mixing integral therefore extends over unphysical regions where $\sigma<0$.
A convenient fix is to truncate the Gaussian mixing distribution at $\sigma=0$ (and renormalise it) as proposed in \citet{bi2}. 
On the other hand, the GHD is based on a non-Gaussian mixing distribution with positive support for $\sigma^2$.
In brief, while \citet{bi2}
use a mixture of (slightly) non-Gaussian distributions with (truncated) Gaussian mixing,
we use a mixture of normals with a strongly non-Gaussian mixing distribution.
 
Our approach offers multiple benefits: 
i) the GHD has long been studied and its mathematical properties are well known;
ii) it has an analytical expression with several different parameterisations;
iii) its moment generating function is analytical and expressions are available
for its first four cumulants;
iv) it reduces to the Gaussian distribution in some limit;
v) optimised techniques have been developed for estimating its parameters given a set of data;
vi) packages in the most popular computer languages are available for its efficient evaluation and also for the estimation of its parameters.

\subsubsection{Cosmology dependence}
\label{sec:cosmo}
\begin{figure}
	\centering
	\includegraphics[scale=0.61]{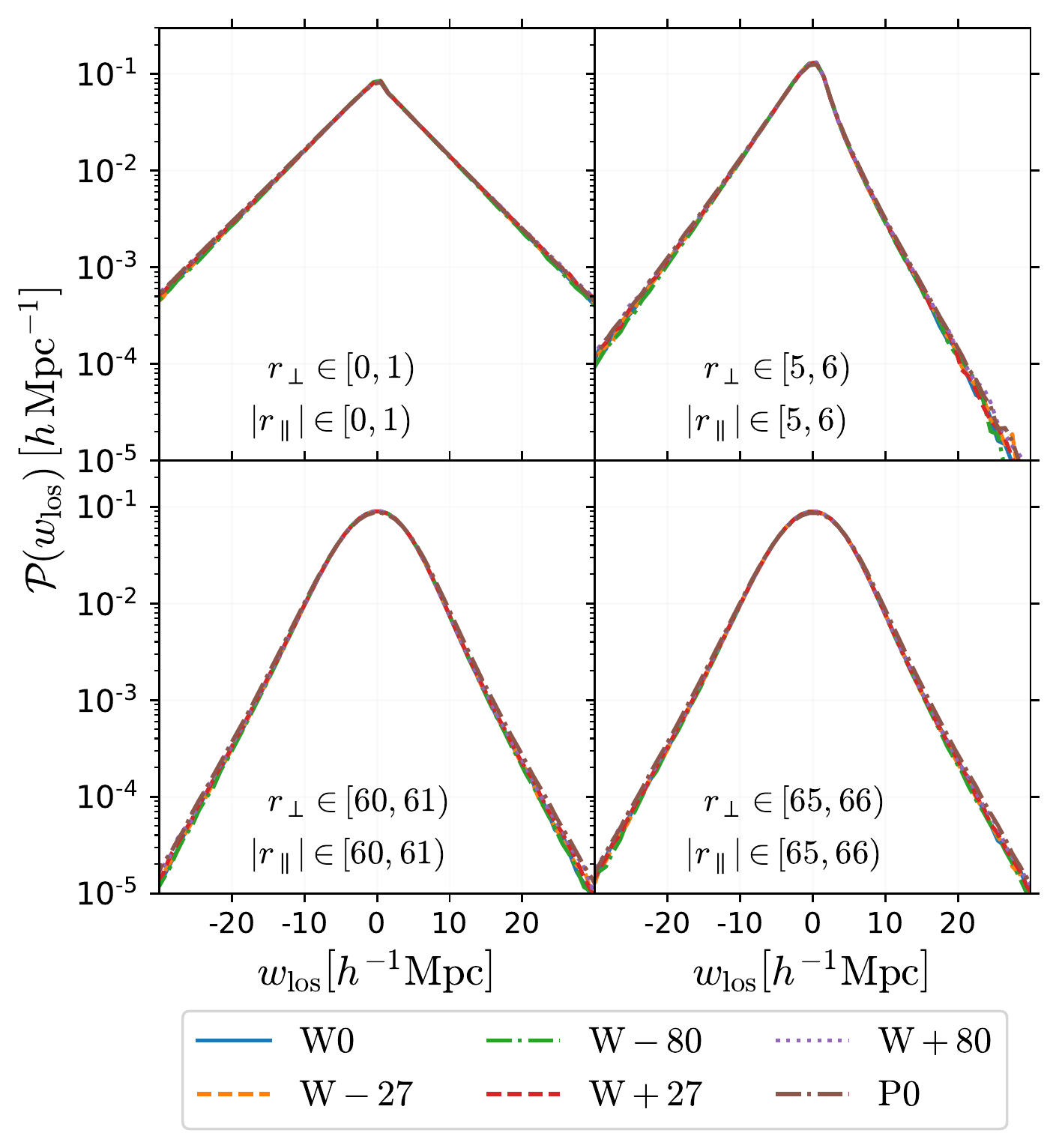}
	\caption{The los pairwise velocity distribution measured at $z=0$ for the DM particles 
in the six cosmological models introduced in
Table~\ref{table:cos}. Spatial separations are given 
in units of $h^{-1} \mathrm{Mpc}$.}
	\label{fig:planckwmap}
\end{figure}

Current cosmic-microwave-background experiments and large-scale-structure studies have
set tight constraints on the cosmological parameters
of the $\Lambda$CDM model.
One open question is whether the pairwise velocity
PDF changes significantly when the underlying cosmological model is varied within the 
currently allowed region of parameter space.
We investigate this issue in Fig.~\ref{fig:planckwmap}
where we compare the pairwise velocity PDFs at $z=0$ for the DM particles extracted from the 6 simulations
introduced in Section \ref{sec:sim}.
Switching from the WMAP to the Planck cosmology
or considering non-Gaussian initial fluctuations
(even at a level that violates current observational limits)
only introduces minimal changes in
$\mathcal{P}_{w_{\rm los}}$ at all separation vectors. Minute differences are noticeable only in the high-velocity tails. 
This is good news as it means that it should be possible to parameterize the scale-dependence of the PDF in a cosmology-independent way so that to facilitate practical applications of the GHD model.

 \begin{figure}
	\centering
	\includegraphics[scale=0.53]{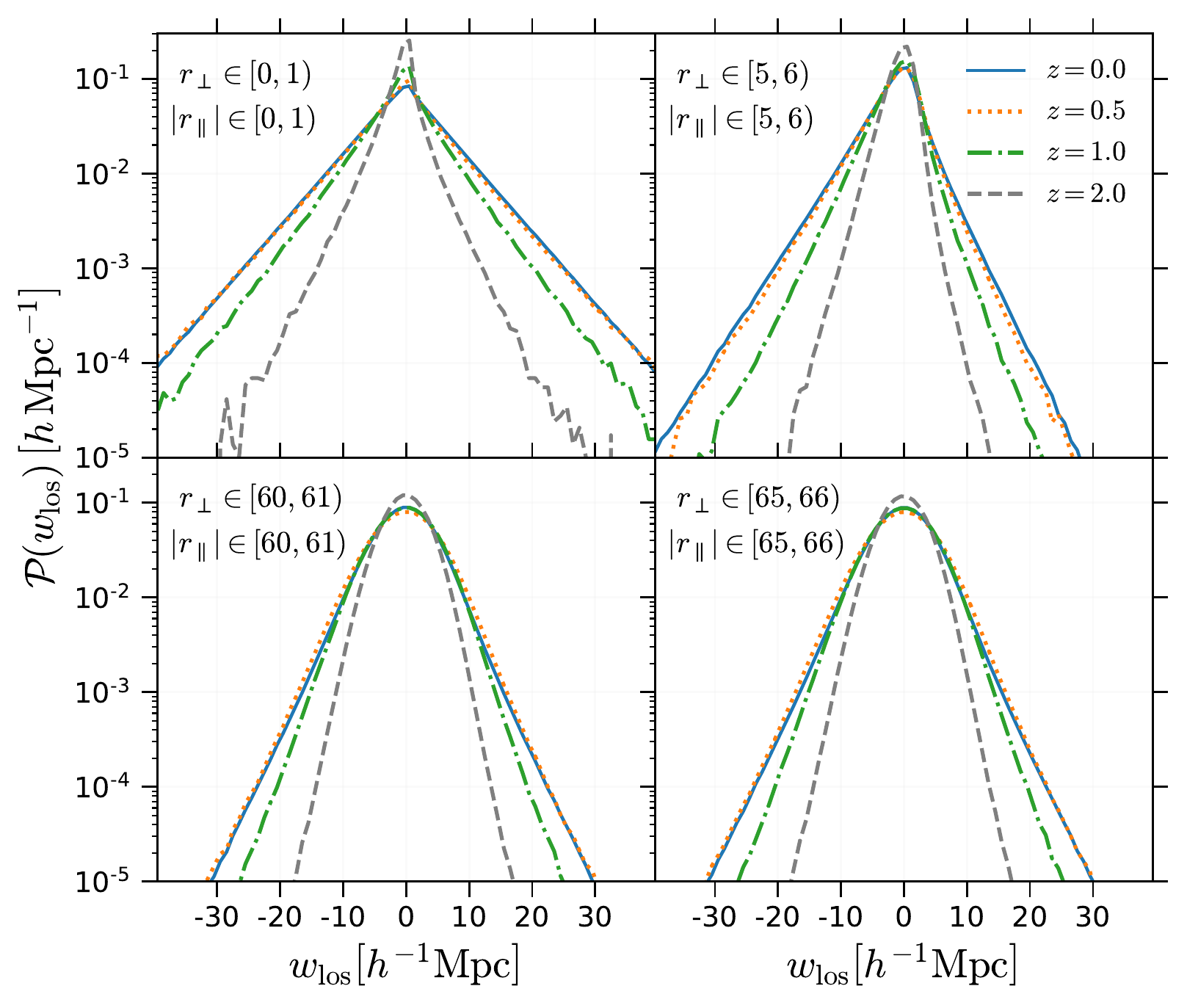}
	\caption{Redshift evolution of     
the los pairwise velocity distribution for the
DM particles of the \textsc{W0} simulation.
Real-space separations are given in units of $h^{-1} \mathrm{Mpc}$.}
	\label{fig:redevo}
\end{figure}

\subsubsection{Redshift evolution}
The pairwise-velocity PDF for the DM particles reflects the non-linear growth of the large-scale structure of
the Universe and, therefore, evolves with time.
This is shown in Fig.~\ref{fig:redevo}, where we
use the W0 simulation to
plot $\mathcal{P}_{w_{\rm los}}$ for various
real-space separations at four different epochs.
At $z=2$, the PDF already exhibits non-Gaussian features including asymmetry around the mode. 
However, the PDF is much more strongly peaked and presents less prominent exponential tails than 
at $z=0$. In fact, only rare massive haloes are resolved at early times and the velocity
PDF is dominated by field-field pairs.
As time goes by, more and more haloes above the mass-resolution limit form and the velocity PDF develops fatter tails.
It is worth stressing that almost no evolution is
noticeable between $z=0.5$ and $z=0$. This is 
a consequence of the fact that the energy budget
of the universe becomes dominated by the cosmological constant and
further development of structure is inhibited. Combined with
the results of Section~\ref{sec:cosmo},
this finding implies that,
for $z\lesssim0.5$, it should be possible to define
a cosmology- and redshift-independent parameterization of the pairwise-velocity PDF, at least for the DM particles.

\subsubsection{DM vs galaxies}

The ultimate application of the streaming model is the interpretation of
the clustering signal extracted for galaxy redshift surveys.
However, our study (like many others before) focusses
on the analysis of simulated DM particles and haloes.
The advantage of using the simulation particles is that they offer a huge statistical sample. On the other hand, the PDF of their pairwise velocities shows stronger exponential tails
with respect to haloes
due to the broadening generated by virial motions and spikier peaks
due to diffuse matter (see Fig.~\ref{fig:rel_los_comp}).
In a sense, a galaxy sample is expected to show intermediate properties between halo and
particle datasets. A pure sample of central galaxies will closely look like a halo sample
while adding more and more satellites will progressively drive the PDF towards the results
for simulation particles.
It is long known that galaxy clustering in redshift space is
well described by the streaming model assuming an exponential $\mathcal{P}_{w_{\rm los}}$
on very small scales and a Gaussian one on very large scales. Since the GHD is of very general form and reduces to these limits for particular combinations of its parameters, we are confident that our analysis can be straightforwardly generalised to galaxies. We will investigate this issue in our future work.

\section{Summary}
\label{sec:summary}
The galaxy 2-point correlation function in redshift space
depends on the orientation of the pair-separation vector with respect to the los.
This anisotropy encodes information about the velocities arising from gravitational
instability. On large scales, where linear perturbation theory applies, RSD 
allow for a measurement of the growth rate of structure.
Combining estimates at different redshifts
help differentiate dark energy models based on General Relativity from modified gravity as the cause of the accelerating Universe.

Modifications to the theory of gravity generally introduce extra degrees of freedom whose
effect (the so-called fifth force) must be suppressed by some screening mechanism on scales where General Relativity is well tested. 
Such constraint implies that characteristic signatures will be imprinted on intermediate cosmological scales.
In fact, the screening mechanism is expected to
affect the non-linear clustering and the velocities of tracers of the large-scale structure.
Testing these predictions provides a strong motivation to extend the analysis of galaxy clustering to smaller scales than currently done in cosmological studies.

Realising this program in practice requires, however, a number of tools.
Among them, robust theoretical predictions for the modified theories of gravity together with
an accurate (and, possibly, non-perturbative) description of RSD.
In this paper, we focussed on the latter issue. In particular, we discussed the classical streaming model for RSD and how its implementation can be improved to get accurate predictions 
at non-linear scales. Our main results can be summarised as follows.

In Section \ref{sec:str},
starting from the one- and two-particle phase-space densities, we derived the governing
equations of the model. For ordered pairs, we obtained equation (\ref{eq:ex_str})  
which coincides with the standard equation discussed in the literature. Our result is exact
and holds true also in the case of multi streaming thanks to our particle-based approach.
The correct solution for unordered pairs has been given in equation (\ref{eq:streaming}).
The modifications with respect to equation (\ref{eq:ex_str}) account for the pairs that
reverse their los ordering between real and redshift space. These swaps occur more frequently for pairs with small spatial separations.

After briefly reviewing the history of the streaming model and of its applications,
we investigated the limitations of using the Gaussian
ansatz for the pairwise-velocity PDF. 
In agreement with previous studies,
we showed that this approximation
fails to reproduce the outcome of $N$-body simulations for redshift-space separations 
$s\lesssim 20\, h^{-1} \textrm{Mpc}$
while it achieves percent level accuracy on $\xi_{\textrm{s}}$ on substantially larger scales.
Our analysis revealed,
however, that a Gaussian PDF
never manages to reproduce the integrand of the streaming equation to the same level of accuracy.
The success of the GSM on large scales, therefore, originates
from fortuitous cancellations between
the contributions of the peak and the wings in the integrand of the streaming equation (see Fig.~\ref{fig:clpt_xis}).

In Section \ref{stats}, we used a high-resolution $N$-body simulation to investigate 
how the PDF of the los component of the pairwise-velocity, $w_\parallel$, depends on the
pair separation vector for both DM particles and haloes. 
The first four cumulants of the PDF
show a complex pattern (Fig.~\ref{fig:mo_wp})
that can be understood in terms of a few
isotropic components and the angle that the pair separation forms with the los.
We derived a general relation between the cumulants of $w_\parallel$ and those
of the radial and transverse pairwise velocities, equation (\ref{cumulants}).
We then studied the scale-dependence
of the isotropic components 
in the simulation (Fig.~\ref{fig:mo_radtan}) and 
measured the pairwise-velocity bias of
the DM haloes (Fig.~\ref{fig:pairwise_bias}).
Additionally, we dissected the
pairwise-velocity PDF for the
DM and
showed that the tails 
are generated by particles
in massive haloes while the region around the mode is dominated by field particles (Figs. \ref{fig:rel_los_comp} and \ref{fig:rel_los_comp_halos}).

Finally, in Section \ref{sec:phe}, we proposed an analytical fitting function for the pairwise-velocity distribution and demonstrated that it provides an excellent description of numerical data.
We first introduced the mathematical background of mixtures and then described the properties
of the GHD, a unimodal PDF with exponential tails.  
Comparing with $N$-body simulations,
we showed that the GHD is able to approximate 
the PDF of the los pairwise velocity
at all scales 
(for both DM particles and haloes) 
with minimal information loss compared to the common exponential and Gaussian fits (Figs.~\ref{fig:gh}, \ref{fig:KL}, and \ref{fig:corr_gh}). The main drawback to using the GHD in future practical applications is that it depends on 5 tunable parameters. In fact, fixing their values to best fit the $N$-body data gives rise to non-trivial scale dependencies (Fig. \ref{fig:GHDparams}). However, the best-fitting values are strongly correlated and tend to populate a lower-dimensional sequence in 5-dimensional parameter space. 
Using principal-component analysis to exploit the correlations, we managed to reduce the complexity of the model while still providing a remarkable fit to $\xi_{\rm s}$.
We found that 2 parameters are enough on large scales ($s>70\,h^{-1}$ Mpc) while at least 3 are needed on smaller scales.
In this case, even the integrand
of the streaming equation is accurately reproduced by the model
(Fig.~\ref{fig:reducedGH}).

Intriguingly, we found that the pairwise-velocity PDF for the DM at $z=0$ shows only minimal changes when the underlying cosmological parameters
are varied within the current constraints for the $\Lambda$CDM model (Fig.~\ref{fig:planckwmap}).
Moreover, it does not show any
noticeable time evolution for
redshifts $z\lesssim 0.5$ 
(Fig.~\ref{fig:redevo}).
All this suggests that, for low-redshift tracers, it should be
possible to find a cosmology- and
redshift-independent parameterization for the PDF of their pairwise velocities as a function of the separation vector. This would
greatly simplify the implementation
of the GHD model for studying 
anisotropic galaxy clustering.
We will investigate the applicability of these findings to forthcoming data in our future work.

	\section*{Acknowledgements}
	
	We acknowledge partial financial support from the Deutsche Forschungsgemeinschaft through the Bonn-Cologne Graduate School for Physics and Astronomy and the Transregio 33 `The Dark Universe'.
    We are thankful  to  the  community for developing  and  maintaining open-source software packages extensively used in our work, namely Cython \citep{cython}, emcee \citep{emcee}, Matplotlib \citep{matplotlib}, Numpy \citep{numpy}, Scikit-learn \citep{scikit-learn} and Scipy \citep{scipy}.

	
	
	
	\bibliographystyle{mnras}
	\bibliography{streaming} 

	
	

\appendix
\section{Reversed pairs}
\label{app:revpairs}
\begin{figure}
		\centering
		\includegraphics[scale=0.7]{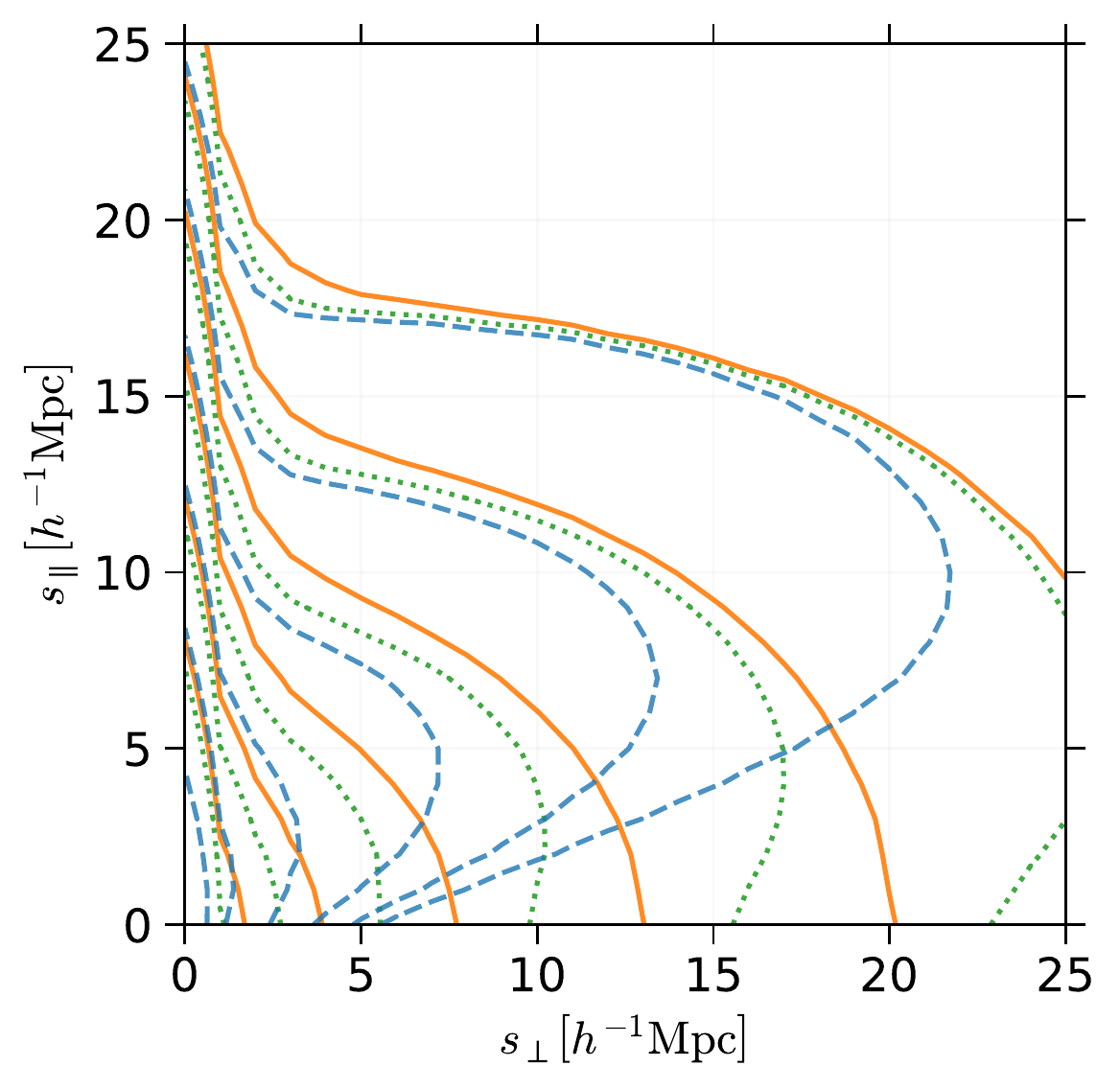}
		\caption{Impact of the reversed pairs on $\xi_{\rm s}$.
        The solid lines show the contour levels obtained using
        equation~(\ref{eq:streaming}) which is exact for ordered pairs. In this case,
        the correlation function nicely matches the measurements presented in Fig.~\ref{fig:mi_corr}. 
                This result is compared with two approximations that do not properly account for the reversed pairs. The dashed lines only considers the region with $r_\parallel>0$ in equation~(\ref{eq:streaming}) and thus completely disregards the swapped pairs. The dotted lines, on the other hand, are obtained by replacing $\mathcal{P}_{w_\parallel}$ with $\mathcal{P}_{w_{\rm los}}$ in equation~(\ref{eq:ex_str}) which improperly weighs the reversed
        pairs.}          
		\label{fig:new_str}
	\end{figure}

In Fig.~\ref{fig:new_str}, we quantify the importance of pair reversals for the two-point
correlation function in redshift space. After measuring $\xi(r)$ and $\mathcal{P}_{w_{\rm los}}$ in our simulation, we apply equation~(\ref{eq:streaming}) and obtain
the solid contour levels for $\xi_s$ that nicely match those shown in the right-hand panel of Fig.~\ref{fig:mi_corr}.
The dashed contours, instead, are computed integrating over the positive values of $r_\parallel$ only, which corresponds to neglecting the reversed pairs.
Not only 
this incomplete model underestimates $\xi_{\rm s}$ by a factor of two
in the FoG regime ($s_\perp<2\,h^{-1}$ Mpc and $s_\parallel \gg
s_\perp$ where reversed pairs are as many as the preserved ones), but also
severely suppresses the redshift-space correlation whenever $s_\parallel \ll s_\perp$. 
The reversed pairs thus give an important contribution on
a vast range of redshift-space separations that extend well beyond the typical size of DM haloes.

\begin{figure}
	\centering
	\includegraphics[scale=0.58]{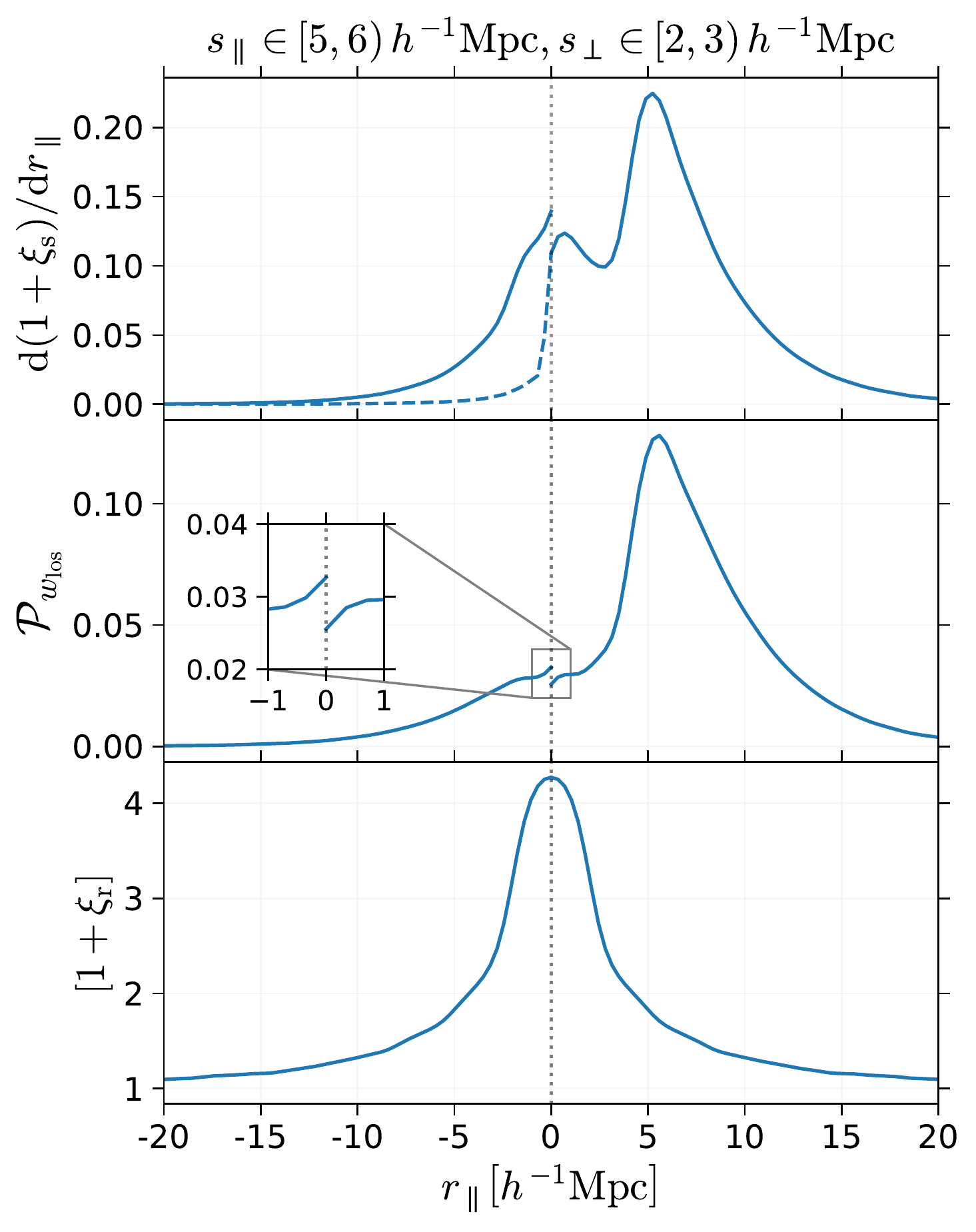}
	\caption{Top: The solid line shows the function $\dif(1+\xi_{\rm s})/\dif r_\parallel$ --
    i.e. the integrand in equation (\ref{eq:streaming}) -- for a narrow range of redshift-space separations indicated above the plot. The dashed line for $r_\parallel<0$ shows the effect of removing the sgn function from the argument of the velocity PDF in equation (\ref{eq:streaming}). Note that this severely miscounts the reversed pairs
    ultimately leading to an underestimate of $\xi_{\rm s}$.
Middle: The contribution to the integrand due to the pairwise-velocity PDF,  
$\mathcal{P}_{w_{\mathrm{los}}}[(s_{\parallel} - r_{\parallel})\,\mathrm{sgn}(r_{\parallel}) \mid \mathbf{r}]$, is plotted as a function of $r_\parallel$.
Bottom: The contribution to the integrand due to real-space clustering, $1+\xi(r)$, is
shown as a function of $r_\parallel$.} 
	\label{fig:consistency1}
	\end{figure}

It is also interesting to explore what would happen if one would naively replace 
$\mathcal{P}_{w_\parallel}$ with $\mathcal{P}_{w_{\rm los}}$ in equation~(\ref{eq:ex_str}) or, equivalently, drop the sgn function in equation~(\ref{eq:streaming}). 
The corresponding result for $\xi_{\rm s}$ is shown with dotted lines in Fig.~\ref{fig:new_str} and it underestimates the actual correlation function on many scales although less severely than in the previous case.  To further investigate the origin of the differences,
in the top panel of Fig.~\ref{fig:consistency1}, we consider a narrow range of redshift-space separations ($s_{\parallel} \in [5,6) \,h^{-1}\mathrm{Mpc}$ and $s_{\perp} \in [2,3)\,h^{-1}\mathrm{Mpc}$) and plot the integrand of equation~(\ref{eq:streaming}) as a function of $r_\parallel$ (solid line). For comparison, we also indicate with a dashed line
the integrand obtained neglecting the sgn function in the argument of $\mathcal{P}_{w_{\mathrm{los}}}$ for $r_\parallel<0$. 
Although the reversed pairs are taken into account in the latter case, they are miscounted which leads to the underestimation of $\xi_{\rm s}$. 

Note that the correct function $\dif(1+\xi_{\rm s})/\dif r_\parallel$ is
discontinous at $r_\parallel=0$. This feature originates from the function
$\mathcal{P}_{w_{\mathrm{los}}}[(s_{\parallel} - r_{\parallel})\,\mathrm{sgn}(r_{\parallel}) \mid \mathbf{r}]$ which is plotted in the middle panel of Fig.~\ref{fig:consistency1}: the factor sgn$(r_\parallel)$ makes sure that
the (continuous but asymmetric) velocity PDF is sampled at $w_{\rm los}=s_\parallel-r_\parallel$ for
$r_\parallel \to 0^+$ and at $w_{\rm los}=r_\parallel-s_\parallel$ for $r_\parallel\to 0^-$.
Also note that $\mathcal{P}_{w_{\mathrm{los}}}$ presents a prominent peak around
$r_{\parallel}\simeq s_{\parallel}$ corresponding to pairs with relatively
small relative los velocity. 
The enhancement of the region around $r_\parallel \sim 0$ in the full integrand
is due to the term $1+\xi(r)$ which is plotted in the bottom panel of the figure.
In simple words, real-space clustering boosts the contribution from close pairs
with large pairwise los velocity.

It is worth mentioning that,
for much larger redshift-space separations (e.g. $s_\parallel \gg 20 \,h^{-1}\mathrm{Mpc}$
and $s_\perp \gg 20 \,h^{-1}\mathrm{Mpc}$), the peak around $r_{\parallel} \approx 0$ is suppressed and the impact of the reversed pairs becomes much less important. 
In this case, neglecting the sgn function in equation~(\ref{eq:streaming}) or using $\mathcal{P}_{w_{\rm los}}$ in equation~(\ref{eq:ex_str}) generates only small errors. 

\color{black}


\bsp	
\label{lastpage}
\end{document}